\documentclass[sigconf,anonymous=false]{acmart}

\AtBeginDocument{%
  \providecommand\BibTeX{{%
    \normalfont B\kern-0.5em{\scshape i\kern-0.25em b}\kern-0.8em\TeX}}}

\setcopyright{acmcopyright}
\copyrightyear{2022}
\acmYear{2022}
\setcopyright{acmcopyright}\acmConference[SIGIR '22]{Proceedings of the 45th
International ACM SIGIR Conference on Research and Development in Information
Retrieval}{July 11--15, 2022}{Madrid, Spain}
\acmBooktitle{Proceedings of the 45th International ACM SIGIR Conference on
Research and Development in Information Retrieval (SIGIR '22), July 11--15, 2022,
Madrid, Spain}
\acmPrice{15.00}
\acmDOI{10.1145/3477495.3531975}
\acmISBN{978-1-4503-8732-3/22/07}

\usepackage{algorithm}
\usepackage{algorithmic}
\usepackage{multirow}
\usepackage{balance}
\usepackage{subfigure}

\usepackage{ragged2e}

\newcommand{\myfont}{\fontsize{8.4pt}{\baselineskip}\selectfont}

\newcommand{\nosection}[1]{\vspace{2pt}\noindent\textbf{#1.}}

\newcommand{\newmodelname}{\textbf{VDEA}}

\begin{document}

\title{Exploiting Variational Domain-Invariant User Embedding for Partially Overlapped Cross Domain Recommendation}

\author{Weiming Liu}
\affiliation{
\institution{College of Computer Science, Zhejiang University}
\country{China}
}
\email{21831010@zju.edu.cn}

\author{Xiaolin Zheng}
\affiliation{
\institution{College of Computer Science, Zhejiang University}
\country{China}
}
\email{xlzheng@zju.edu.cn}

\author{Jiajie Su}
\affiliation{
\institution{College of Computer Science, Zhejiang University}
\country{China}
}
\email{jiajiesu@zju.edu.cn}

\author{Mengling Hu}
\affiliation{
\institution{College of Computer Science, Zhejiang University}
\country{China}
}
\email{humengling@zju.edu.cn}

\author{Yanchao Tan}
\affiliation{
\institution{College of Computer Science, Zhejiang University}
\country{China}
}
\email{yctan@zju.edu.cn}

\author{Chaochao Chen}
\authornote{Chaochao Chen is the corresponding author.}
\affiliation{
\institution{College of Computer Science, Zhejiang University}
\country{China}
}
\email{zjuccc@zju.edu.cn}


\begin{abstract}
Cross-Domain Recommendation (CDR) has been popularly studied to utilize different domain knowledge to solve the cold-start problem in recommender systems.
Most of the existing CDR models assume that both the source and target domains share the same overlapped user set for knowledge transfer.
However, only few proportion of users simultaneously activate on both the source and target domains in practical CDR tasks.
In this paper, we focus on the \textit{Partially Overlapped Cross-Domain Recommendation} (\textit{POCDR}) problem, that is, how to leverage the information of both the overlapped and non-overlapped users to improve recommendation performance. 
Existing approaches cannot fully utilize the useful knowledge behind the non-overlapped users across domains, which limits the model performance when the majority of users turn out to be non-overlapped.
%
%
To address this issue, we propose an end-to-end dual-autoencoder with Variational Domain-invariant Embedding Alignment (\textbf{\newmodelname}) model, a cross-domain recommendation framework for the POCDR problem, which utilizes dual variational autoencoders with both local and global embedding alignment for exploiting domain-invariant user embedding.
\textbf{\newmodelname} first adopts variational inference to capture collaborative user preferences,
and then utilizes Gromov-Wasserstein distribution co-clustering optimal transport to cluster the users with similar rating interaction behaviors.
Our empirical studies on Douban and Amazon datasets demonstrate that \newmodelname~significantly outperforms the state-of-the-art models, especially under the POCDR setting.
\end{abstract}



\maketitle

\setlength{\floatsep}{4pt plus 4pt minus 1pt}
\setlength{\textfloatsep}{4pt plus 2pt minus 2pt}
\setlength{\intextsep}{4pt plus 2pt minus 2pt}
\setlength{\dbltextfloatsep}{3pt plus 2pt minus 1pt}
\setlength{\dblfloatsep}{3pt plus 2pt minus 1pt}
\setlength{\abovecaptionskip}{3pt}
\setlength{\belowcaptionskip}{2pt}
\setlength{\abovedisplayskip}{2pt plus 1pt minus 1pt}
\setlength{\belowdisplayskip}{2pt plus 1pt minus 1pt}

\section{Introduction}








\begin{figure}[t]
\centering
\includegraphics[width=\linewidth]{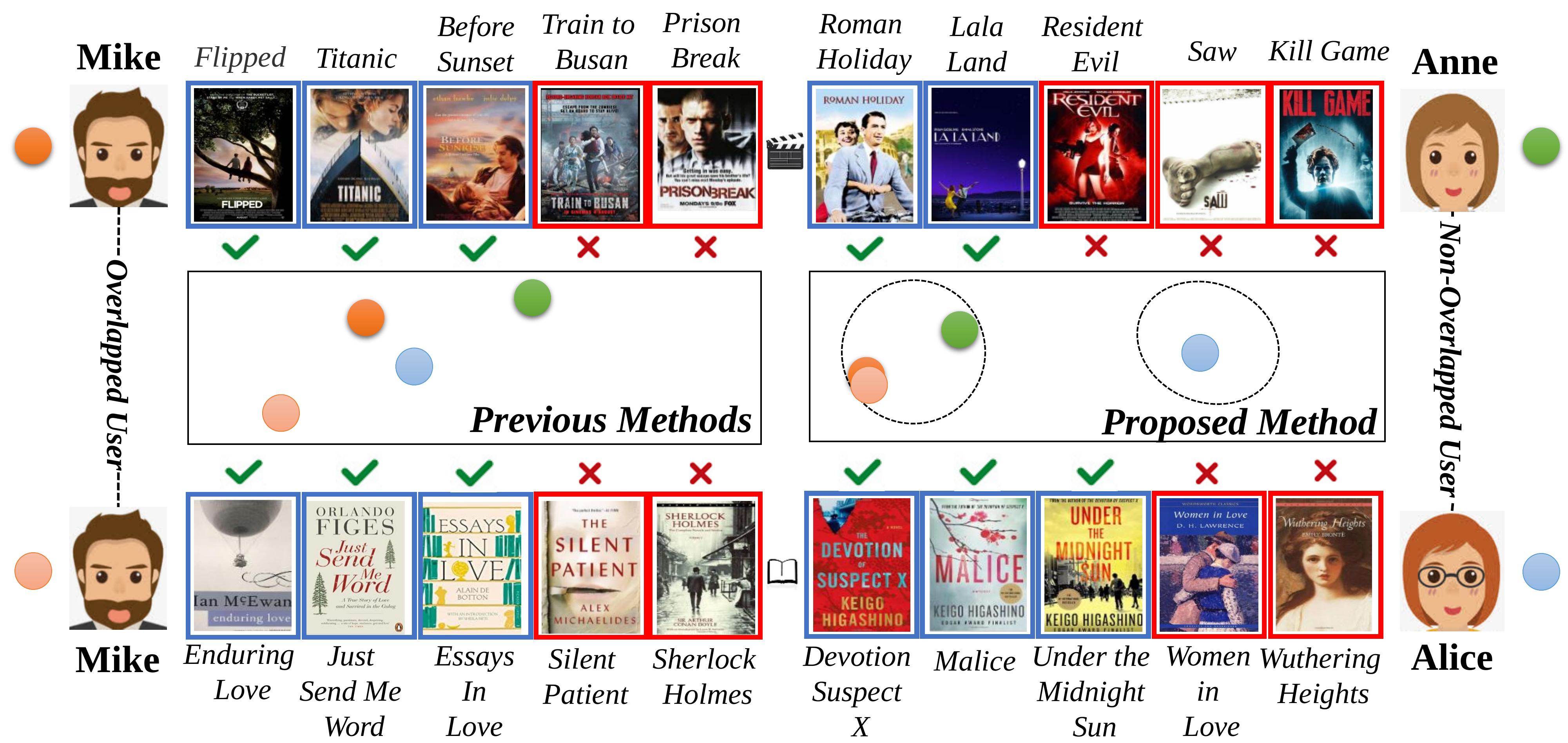}
\caption{The latent user embedding should be domain-invariant, capturing the main characteristics of the users across different domains. For example, Mike and Anne who both like romantic items and dislike horror items should have similar domain-invariant embeddings. In contrast, Mike and Alice have different preferences, and thus their embeddings should be pushed away. }
\label{fig:toy_example}
\end{figure}

Recommender systems become more and more indispensable in practice since it is a sharp tool for users to easily find what they desire. 
Nowadays, users always participate in multiple domains (platforms) for different purposes, e.g., buying books on \emph{Amazon} and watching movies on \emph{Youtube}.
Therefore, Cross Domain Recommendation (CDR) has emerged to utilize both the source and target information for establishing high accuracy recommendation system and overcoming the long-standing data sparsity problem.
Most existing CDR models assume the existence of fully overlapped users or items, who have similar tastes or attributions \cite{cdrbook}, across domains. 
For example, a user who likes romantic movies and dislikes horror movies would always like romantic books and dislike horror books as well.
Although the movie and book domains are rather different in the specific form of expression, the latent user embedding should properly represent the domain-invariant user personal characteristics such as preferring romantic items and disliking horror items, as shown in Fig.~1.
Therefore, existing CDR models can leverage the information from a (source) domain to improve the recommendation accuracy of other (target) domains \cite{cdrsurvey,tanzong}.
%
%
However, in practice, not all users or items are strictly overlapped in CDR settings, which limits the applications of existing CDR approaches.

In this paper, we focus on a general problem in CDR based on our observation. 
That is, two or multiple domains only share a certain degree of overlapped users or items.
Specifically, we term this problem as \textit{Partially Overlapped Cross Domain Recommendation} (\textit{POCDR}). 
On the one hand, POCDR has some overlapped users who have rating interactions in both the source and target domains, similar as the conventional CDR problem. 
On the other hand, POCDR also has several non-overlapped users who only have rating interactions in at most one domain, making it different from the conventional CDR problem.
Fig.~2 shows an example of the POCDR problem\footnote{Note that the examples and our proposed models in this paper could be easily adapted to the situations where the source and target domains share partially overlapped items. }, where Jimmy and Mary are overlapped users, while John, Tom, Lily, and Rosa are non-overlapped users.
%
%
The prime task of POCDR is to solve the data sparsity problem by leveraging the user-item interactions, with only partially overlapped users across the source and target domains.
Solving POCDR is challenging, mainly because it is hard to fully exploit and transfer useful knowledge across domains, especially when extremely few users are overlapped.

Existing researches on CDR cannot solve the POCDR problem well. 
Firstly, conventional CDR approaches can only adjust to the circumstances where the majority of users are overlapped, but they cannot maintain the prediction performance when the majority of users turn out to be non-overlapped.
The overlapped users can be utilized as a bridge to transfer knowledge across domains, while knowledge of the non-overlapped users cannot be extracted efficiently, which leads to unreliable predictions. 
Secondly, there exist non-overlapped CDR models which could adopt group-level codebook transfer to provide recommendation \cite{cbt,cbt2}.
However, they cannot ensure that the shared embedding information across domains is consistent, which limits the effectiveness of knowledge transfer \cite{kerkt}. 
%
%
Take the toy example shown in the left of Fig.~\ref{fig:toy_example} for example, lacking effective constraint on the non-overlapped users leads to the consequence that Alice who likes horror but dislikes romantic has the nearest embedding distance with Mike who likes romantic but dislikes horror instead. 
Obviously, insufficient training of the non-overlapped users finally will deteriorate the model performance.
Meanwhile, conventional CDR approaches have concluded that clustering the overlapped users with similar characteristics can also promote the model performance \cite{cluster1,cluster2}. 
However, how to properly co-cluster similar overlapped and non-overlapped users across domains is still a serious challenge in POCDR setting.
In a nutshell, current CDR models cannot achieve satisfying performance because of the lacking of sufficient techniques to fully leverage the data in both domains under the POCDR setting.

To address the aforementioned issue, in this paper, we propose \newmodelname, an end-to-end dual-autoencoder with Variational Domain-invariant Embedding Alignment (\textbf{\newmodelname}), for solving the POCDR problem. 
%
%
The core insight of \newmodelname~is exploiting domain-invariant embedding across different domains, for both the overlapped and non-overlapped users. 
Again, in the example of Fig.\ref{fig:toy_example}, Mike and Anne who share similar tastes, i.e., preferring romantic items rather than horror ones, should have similar domain-invariant embeddings.
In contrast, users with different tastes, e.g., Mike and Alice, should be pushed away in the latent embedding space.
In order to better represent user embedding with high-quality  rating prediction and knowledge transfer ability, we utilize two modules in \newmodelname, i.e., \textit{variational rating reconstruction module} and \textit{variational embedding alignment module}, as will be shown in Fig.~3. 
On the one hand, the variational rating reconstruction module aims to capture user collaborative preferences in the source and target domains with cross-entropy loss.
On the other hand, the variational rating reconstruction module can cluster users with similar tastes or characteristics on Mixture-Of-Gaussian distribution simultaneously.
The variational embedding alignment module further includes two main components, e.g, \textit{variational local embedding alignment} and \textit{variational global embedding alignment}.
Variational local embedding alignment tends to match the overlapped users across domains to obtain more expressive user embeddings.
Variational global embedding alignment can co-cluster the similar user groups via distribution optimal transport, transferring useful knowledge among both overlapped and non-overlapped users.
By doing this, we can not only obtain discriminative domain-invariant user embedding in a single domain, but also cluster users with similar preferences between the source and target domains. 

We summarize our main contributions as follows:
(1) We propose a novel framework, i.e., \newmodelname, for the POCDR problem, which can utilize the rating interactions of both overlapped and non-overlapped users to improve the recommendation performance.
(2) In order to better exploit users with similar tastes or characteristics, we equip \newmodelname~with variational inference and distribution co-clustering optimal transport to obtain more discriminative and domain-invariant latent user embedding. 
(3) Extensive empirical studies on Douban and Amazon datasets demonstrate that \newmodelname~significantly improves the state-of-the-art models, especially under the POCDR setting.

\section{Related Work}

\nosection{Cross Domain Recommendation}
Cross Domain Recommendation (CDR) emerges as a technique to alleviate the long-standing data sparsity problem in recommendation by assuming the same user set across domains \cite{cbt,tr2013,liuwww, tanicde}. 
According to \cite{cdrsurvey}, existing CDR models have three main types, i.e., \textit{transfer-based} methods, \textit{clustered-based} methods, and \textit{multitask-based} methods.
Transfer-based methods \cite{catn,emcdr} learn a linear or nonlinear mapping function across domains.
Some recent method \cite{etl} even adopts equivalent transformation to obtain more reliable knowledge across domains with shared users. 
Clustered-based methods \cite{cdiec} adopt the co-clustering approach to learn cross-domain comprehensive embeddings by collectively leveraging single-domain and cross-domain information within a unified framework.
Multi-task-based methods \cite{conet,dtcdr,gadtcdr} enable dual knowledge transfer across domains by introducing shared connection modules in neural networks.
However, the majority of current CDR frameworks have the underlying assumption that the source and target domains share the same set of users or items \cite{cdrsurvey}.
Conventional CDR approaches cannot perfectly solve the POCDR problem where a large number of users are non-overlapped.  
As a comparison, in this paper, we focus on the POCDR problem, and propose \newmodelname~which adopt variational local and global embedding alignment for better exploiting domain-invariant embeddings for all the users across domains.

\begin{figure}
\centering
\includegraphics[width=\linewidth]{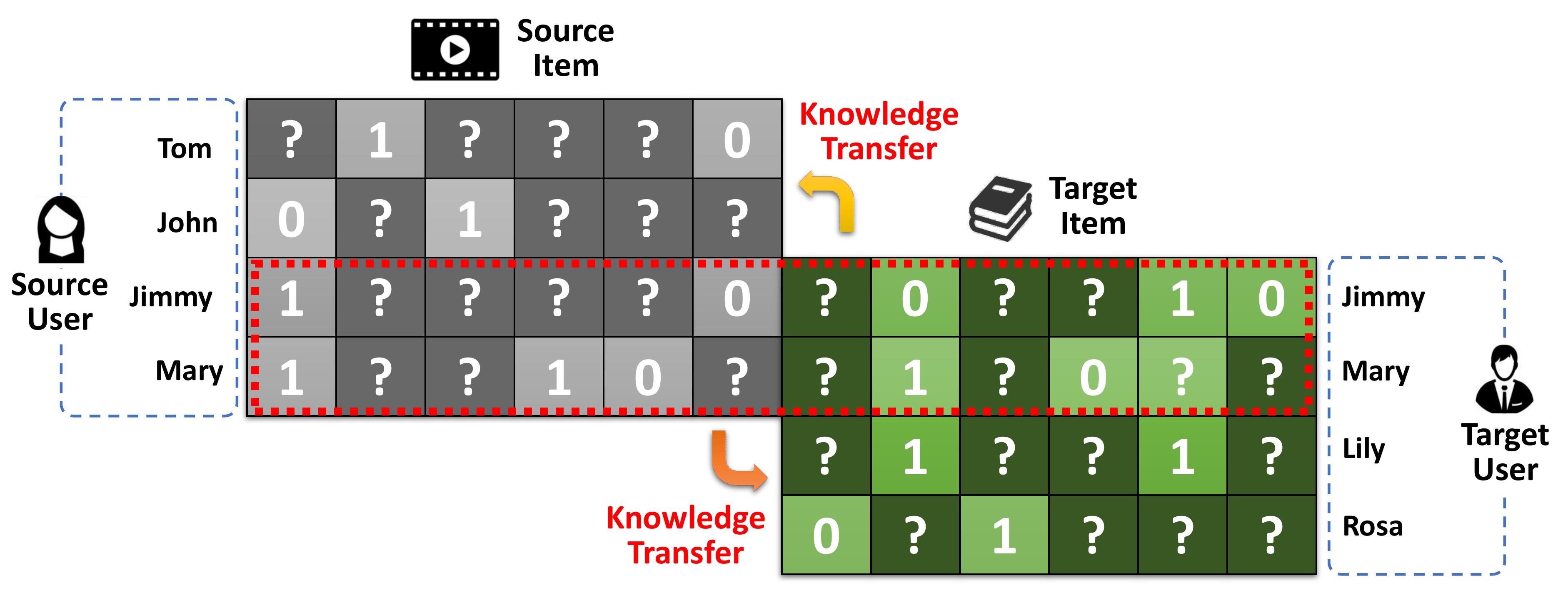}
\caption{The POCDR problem. }
\label{fig:model}
\end{figure}

\nosection{Domain Adaptation}
Domain adaptation aims to transfer useful knowledge from the source samples with labels to target samples without labels for enhancing the target performance.
Existing work on this is mainly of three types, i.e., \textit{discrepancy-based} methods \cite{tca, mmd}, \textit{adversarial-based} methods \cite{dann,cdan}, and \textit{reconstruction-based} methods \cite{ieee_survey,transfer_survey}. 
Discrepancy-based methods, e.g., Maximum Mean Discrepancy (MMD) \cite{mmd}, Correlation Alignment (CORAL). 
Recently, ESAM \cite{esam} extends CORAL with attribution alignment for solving the long-tailed item recommendation problem.
Adversarial-based methods integrate a domain discriminator for adversarial training, including Domain Adversarial Neural Network (DANN) \cite{dann} and Adversarial Discriminative Domain Adaptation (ADDA) \cite{adda}.
DARec \cite{darec} even transferred knowledge across domains with shared users via deep adversarial adaptation technique.
Reconstruction based methods utilize autoencoder to exploit the domain-invariant latent representations, e.g., Unsupervised Image-to-Image Translation Networks (UNIT) \cite{unit,munit}, which indicates that exploiting shared latent space across different domains can enhance the model performance and generalization.
However, most existing reconstruction-based models fail to adopt a dual discriminator network to reduce the distribution discrepancy across different domains.
In this paper, we propose to combine reconstruction-based model with discrepancy-based method for better exploiting domain-invariant embeddings.

\section{Modeling for \newmodelname}

\begin{figure}
\centering
\includegraphics[width=\linewidth]{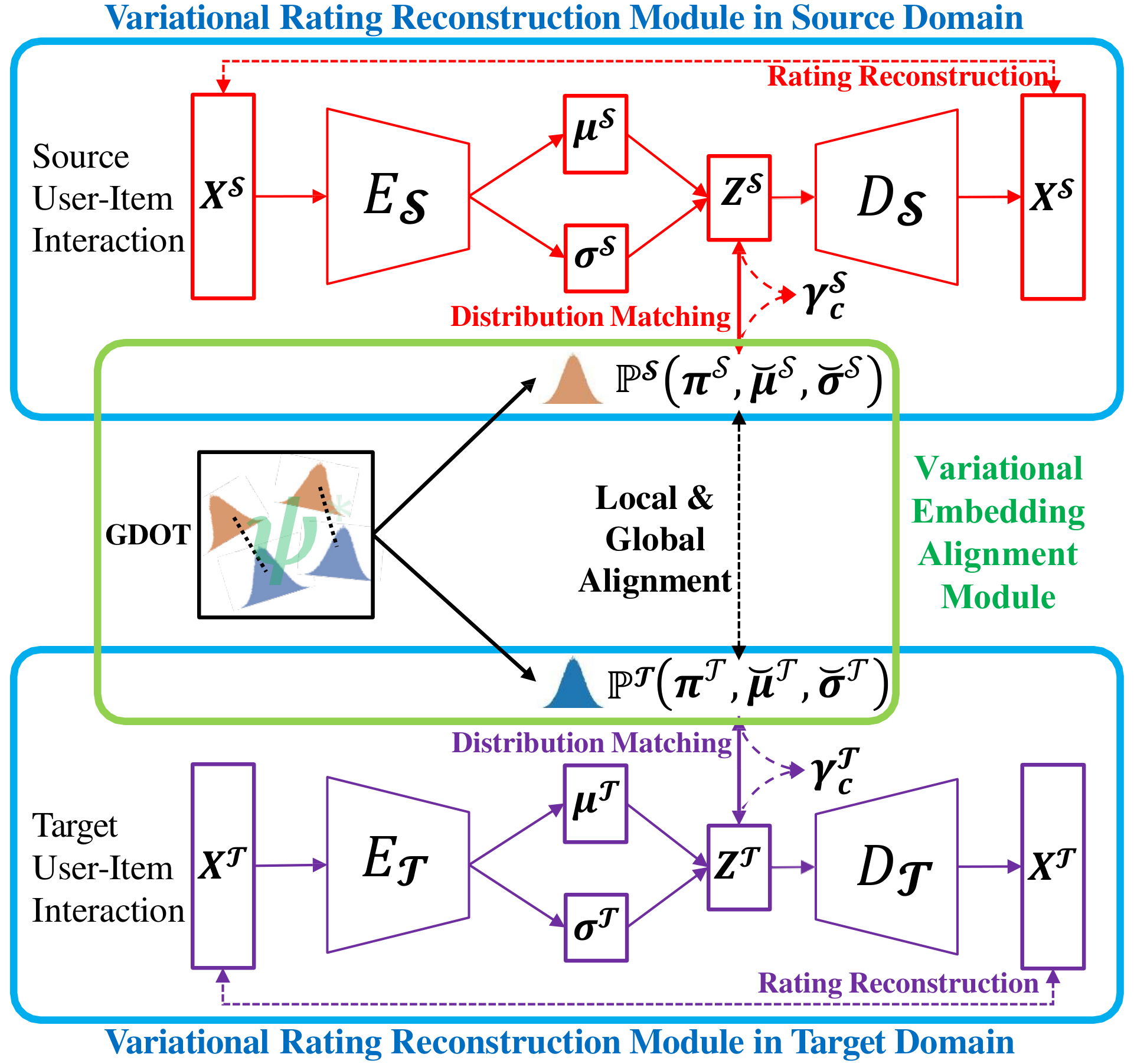}
\caption{The basic framework of \newmodelname. The blue part is the variational rating reconstruction module and the green part is the variational embedding alignment module.}
\vspace{-0.3cm} 
\label{fig:vade}
\end{figure}

\subsection{Framework of \newmodelname}

First, we describe notations.
We assume there are two domains, i.e., a source domain $\mathcal{S}$ and a target domain $\mathcal{T}$.
We assume $\mathcal{S}$ has $|\boldsymbol{U}^\mathcal{S}|$ users and $|\boldsymbol{V}^\mathcal{S}|$ items, and $\mathcal{T}$ has $|\boldsymbol{U}^\mathcal{T}|$ users and $|\boldsymbol{V}^\mathcal{T}|$ items. 
Unlike conventional cross domain recommendation, in POCDR, the source and target user sets are not completely the same.
We use the overlapped user ratio $\mathcal{K}_u \in (0,1)$ to measure how many users are concurrence.
Bigger $\mathcal{K}_u$ means more overlapped users across domains, while smaller $\mathcal{K}_u$ means the opposite situation.

Then, we introduce the overview of our proposed \newmodelname~framework.
The aim of \newmodelname~is providing better single domain recommendation performance by leveraging useful cross domain knowledge for both overlapped and non-overlapped users.
\newmodelname~model mainly has two modules, i.e., \textit{variational rating reconstruction module} and \textit{variational embedding alignment module}, as shown in Fig.~\ref{fig:vade}.
The variational rating reconstruction module establishes user embedding according to the user-item interactions on the source and target domains.
Meanwhile, the variational rating reconstruction module also clusters the users with similar characteristics or preferences based on their corresponding latent embeddings to enhance the model generalization. 
The basic intuition is that users with similar behaviors should have close embedding distances through this module with deep unsupervised clustering approach \cite{cluster1,mvae,cdiec}.
The variational embedding alignment module has two parts, i.e., \textit{variational local embedding alignment} and \textit{variational global embedding alignment}, sufficiently aligning the local and global user embeddings across domains.
The local embedding alignment can utilize the user-item ratings of the overlapped users to obtain more expressive user embeddings \cite{hoff}.
Intuitively, expressive overlapped user embedding can enhance model training of the non-overlapped users, since the non-overlapped users share the similar preference (e.g., romantic or realistic) as the overlapped ones \cite{kerkt}.
Moreover, the global embedding alignment can also directly co-cluster both the overlapped and non-overlapped users with similar tastes, which can boost the procedure of knowledge transfer. 
To this end, the variational embedding alignment module can exploit domain-invariant latent embeddings for both overlapped and non-overlapped users. 
The framework of \newmodelname~is demonstrated in Fig.~\ref{fig:vade} and we will introduce its two modules in details later.

\begin{figure}
\centering
\includegraphics[width=\linewidth]{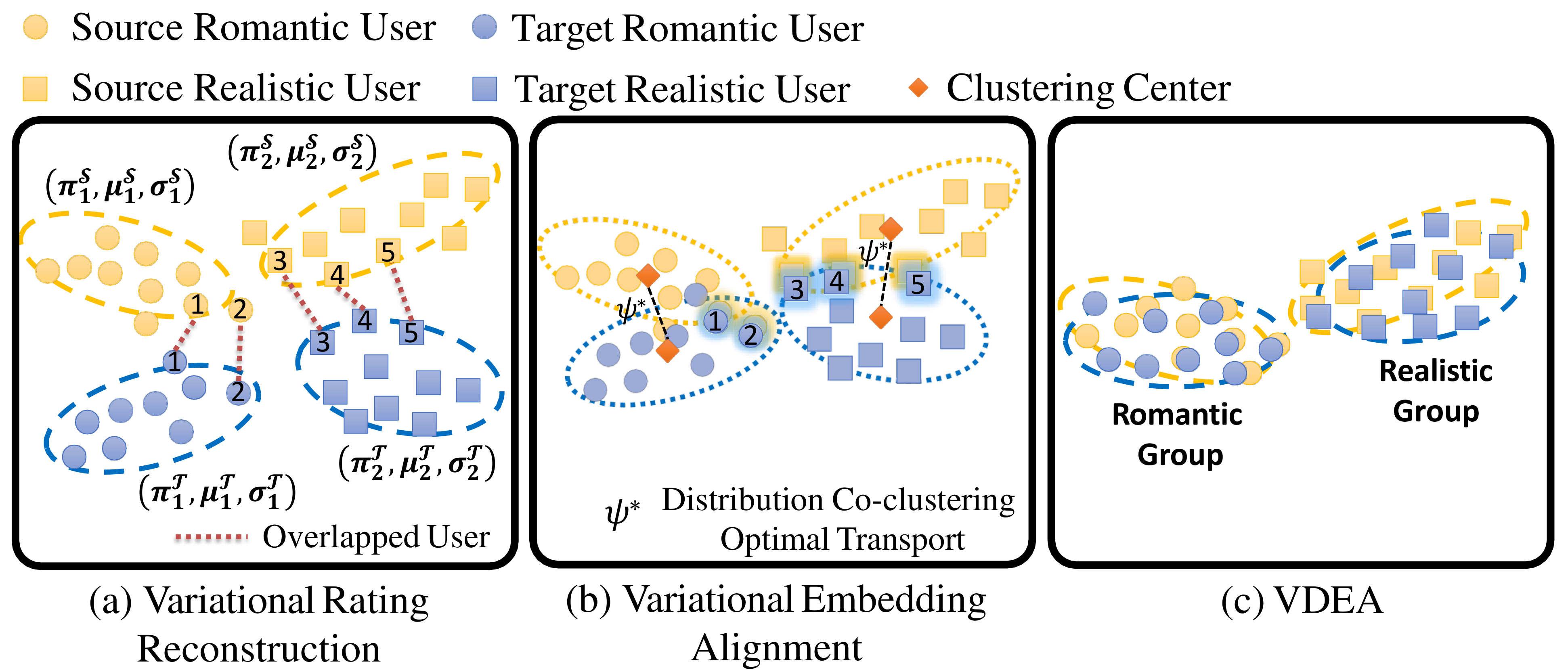}
\caption{Demonstration of the general process of \newmodelname. Notably that the preference attributions (Romantic or Realistic) are just examples and they are not accessible during the training process.}

\label{fig:insight}
\vspace{-0.3cm}
\end{figure}

\subsection{Variational Rating Reconstruction Module}
To begin with, we first introduce the variational rating reconstruction module.
We define $\boldsymbol{X}^\mathcal{S} \in \mathbb{R}^{N \times |\boldsymbol{V}^\mathcal{S}|}$ and $\boldsymbol{X}^\mathcal{T} \in \mathbb{R}^{N \times |\boldsymbol{V}^\mathcal{T}|}$ as the user-item interactions on the source and target domains respectively, and $N$ is batch size.
We adopt dual variational autoencoder as the main architecture for accurately representing and clustering user embedding according to the rating interactions.
We utilize two variational autoenocers in \newmodelname, i.e., a \textit{source variational autoencoder} and a \textit{target variational autoencoder}, for source and target domains, respectively. 
The source variational autoencoder includes the encoder $E_\mathcal{S}$ and decoder $D_\mathcal{S}$. Similarly, the target variational autoencoder includes the encoder $E_\mathcal{T}$ and decoder $D_\mathcal{T}$.
The encoders $E_\mathcal{S}$ and $E_\mathcal{T}$ can predict the mean ($\boldsymbol{\mu}^\mathcal{S}$ and $\boldsymbol{\mu}^\mathcal{T}$) and log variance ($\log (\boldsymbol{\sigma}^\mathcal{S})^2$ and $\log (\boldsymbol{\sigma}^\mathcal{T})^2$) for the input data as $[\boldsymbol{\mu}^\mathcal{S},\log (\boldsymbol{\sigma}^\mathcal{S})^2] = E_\mathcal{S}(\boldsymbol{X}^\mathcal{S})$ and $ [\boldsymbol{\mu}^\mathcal{T},\log (\boldsymbol{\sigma}^\mathcal{T})^2] = E_\mathcal{T}(\boldsymbol{X}^\mathcal{T})$.
In order to guarantee gradient descent in back propagation, we also apply the reparamaterize mechanism \cite{multvae,recvae} as $\boldsymbol{Z}^\mathcal{S} = \boldsymbol{\mu}^\mathcal{S} + \epsilon^\mathcal{S}\boldsymbol{\sigma}^\mathcal{S}$ and $\boldsymbol{Z}^\mathcal{T} = \boldsymbol{\mu}^\mathcal{T} + \epsilon^\mathcal{T}\boldsymbol{\sigma}^\mathcal{T}$, where $\epsilon^\mathcal{S},\epsilon^\mathcal{T} \sim \mathcal{N}(0,I)$ are randomly sampled from normal distribution, and $\boldsymbol{\mu}^\mathcal{S}, \boldsymbol{\mu}^\mathcal{T},\boldsymbol{\sigma}^\mathcal{S}$, and $\boldsymbol{\sigma}^\mathcal{T} \in \mathbb{R}^{N \times D}$ have the same dimension $D$.
Inspired by variational deep embedding \cite{vade}, we expect the latent user embedding conform to Mixture-Of-Gaussian distribution for grouping similar users.
The dual variational autoencoder also reconstructs the user-item interaction as $\hat{\boldsymbol{X}}_\mathcal{S} = D_\mathcal{S}(\boldsymbol{Z}^\mathcal{S})$ and $\hat{\boldsymbol{X}}_\mathcal{T} = D_\mathcal{T}(\boldsymbol{Z}^\mathcal{T})$.
In this section, we will first present the variational evidence lower bound according to the Mixture-Of-Gaussian distributions, and then illustrate the rating reconstruction and distribution matching for the total loss.
By proposing the variational rating reconstruction module, the users with similar characteristics will be clustered in each domain, as can be seen in Fig.~\ref{fig:insight}(a).

\nosection{Variational evidence lower bound}
Since we tend to cluster the users with similar preferences, we choose Mixture-Of-Gaussian as the user latent distribution.
To do this, we let user embedding obey the Mixture-Of-Gaussian distribution and properly predict the user-item rating interactions through variational evidence lower bound \cite{vade}. 
The variational evidence lower bound for the variational rating reconstruction module is to maximize the likelihood of the given user embeddings in Mixture-Of-Gaussian distribution. 
%
%
%
Specifically, we define the prior source and target latent distributions as $\mathbb{P}^\mathcal{S} = {\rm GMM} ({\pi}^\mathcal{S},\breve{\boldsymbol{\mu}}^\mathcal{S},(\breve{\boldsymbol{\sigma}}^\mathcal{S})^2 )$ and $\mathbb{P}^\mathcal{T} = {\rm GMM} ({\pi}^\mathcal{T},\breve{\boldsymbol{\mu}}^\mathcal{T},(\breve{\boldsymbol{\sigma}}^\mathcal{T})^2 )$ respectively, where ${\rm GMM}(\cdot,\cdot,\cdot)$ denotes the distribution of Mixture-Of-Gaussian. 
$\boldsymbol{\pi}^\mathcal{S}$ and $\boldsymbol{\pi}^\mathcal{T}$ denote the prior probability of each cluster components in Mixture-Of-Gaussian,
$(\breve{\boldsymbol{\mu}}^\mathcal{S},(\breve{\boldsymbol{\sigma}}^\mathcal{S})^2 )$ and $(\breve{\boldsymbol{\mu}}^\mathcal{T},(\breve{\boldsymbol{\sigma}}^\mathcal{T})^2 )$ denote the mean and variance of the Gaussian distribution in the source domain and target domain, respectively. 
For the $i$-th user, $\sum_{k=1}^K \boldsymbol{\pi}^\mathcal{S}_{ik} = \boldsymbol{\pi}^\mathcal{T}_{ik} = 1$ with $K$ denoting the total number of clusters.
Taking the joint probability in the source domain for example, it can be factorized as $p_\mathcal{S}(\boldsymbol{X}^\mathcal{S},\boldsymbol{Z}^\mathcal{S},C^\mathcal{S})$ in the perspective of the generative process:
\begin{equation}
\begin{aligned}
\label{equ:elbo_p}
&p_\mathcal{S}(\boldsymbol{X}^\mathcal{S},\boldsymbol{Z}^\mathcal{S},C^\mathcal{S}) = p_\mathcal{S}  (\boldsymbol{X}^\mathcal{S}|\boldsymbol{Z}^\mathcal{S} )p_\mathcal{S} (\boldsymbol{Z}^\mathcal{S}|C^\mathcal{S} )p_\mathcal{S} (C^\mathcal{S} )\\
&p_\mathcal{S} (C^\mathcal{S} ) = {\rm Cat} (C^\mathcal{S} | \boldsymbol{\pi}^\mathcal{S} ) \quad p_\mathcal{S} (\boldsymbol{Z}^\mathcal{S}|C^\mathcal{S} ) = \mathcal{N} (\boldsymbol{Z}^\mathcal{S}|\breve{\boldsymbol{\mu}}^\mathcal{S}_c, (\breve{\boldsymbol{\sigma}}^\mathcal{S}_c )^2\boldsymbol{I} )\\
& p_\mathcal{S}  (\boldsymbol{X}^\mathcal{S}|\boldsymbol{Z}^\mathcal{S} )= {\rm Ber}(\boldsymbol{X}^\mathcal{S}| \boldsymbol{\mu}^\mathcal{S}_x),
\end{aligned}
\end{equation}
where $\boldsymbol{X}^\mathcal{S}$ is the observed sample and $\boldsymbol{Z}^\mathcal{S}$ is the latent user embedding.
The corresponding cluster is $C^\mathcal{S} \sim {\rm Cat}(\boldsymbol{\pi}^\mathcal{S})$ with ${\rm Cat}(\cdot)$ denoting the prior categorical distribution.
$\breve{\boldsymbol{\mu}}^\mathcal{S}_c$ and $(\breve{\boldsymbol{\sigma}}^\mathcal{S}_c)^2$ denote the mean and variance of the Gaussian distribution to the $c$-th cluster.
${\rm Ber}(\boldsymbol{X}^\mathcal{S}| \boldsymbol{\mu}^\mathcal{S}_x)$ denotes multivariate Bernoulli distribution for the output of the decoder.
%

We then adopt Jensen's inequality with the mean-field assumption to estimate the Evidence Lower Bound (ELBO) by maximizing $\log p_\mathcal{S} (\boldsymbol{X}^\mathcal{S})$. 
%
%
Following \cite{vade}, ELBO can be rewritten in a simplified form as:
\begin{equation}
\begin{aligned}
\label{equ:elbos}
&\mathcal{L}_{ELBO}(\boldsymbol{X}^\mathcal{S}) = \mathbb{E}_{q_\mathcal{S} (\boldsymbol{Z}^\mathcal{S},C^\mathcal{S}|\boldsymbol{X}^\mathcal{S})} \left[ \log \frac{p_\mathcal{S}(\boldsymbol{Z}^\mathcal{S},C^\mathcal{S},\boldsymbol{X}^\mathcal{S})  }{q_\mathcal{S} (\boldsymbol{Z}^\mathcal{S},C^\mathcal{S}|\boldsymbol{X}^\mathcal{S})}  \right]\\
& = \mathbb{E}_{q_\mathcal{S} (\boldsymbol{Z}^\mathcal{S},C^\mathcal{S}|\boldsymbol{X}^\mathcal{S})} \left[ \log p_\mathcal{S} (\boldsymbol{X}^\mathcal{S}|\boldsymbol{Z}^\mathcal{S}) + \log \frac{p_\mathcal{S} (\boldsymbol{Z}^\mathcal{S},\boldsymbol{C}^\mathcal{S})}{q_\mathcal{S} (\boldsymbol{Z}^\mathcal{S},C^\mathcal{S}|\boldsymbol{X}^\mathcal{S})}  \right], 
\end{aligned}
\end{equation}
where the first term represents the rating reconstruction loss that encourages the model to fit the user-item interactions well.
The second term calculates the Kullback-Leibler divergence between Mixture-of-Gaussians prior $q_\mathcal{S}(\boldsymbol{Z}^\mathcal{S},C^\mathcal{S}|\boldsymbol{X}^\mathcal{S})$ and the variational posterior $p_\mathcal{S} (\boldsymbol{Z}^\mathcal{S},\boldsymbol{C}^\mathcal{S} )$. 
Furthermore, in order to balance the reconstruction capacity (the first term) efficiently and disentangle the latent factor (the second term) automatically, 
we change the KL divergence into the following optimization equation with constraints:
\begin{equation}
\begin{aligned}
\label{equ:opt}
&\max \mathbb{E}_{q_\mathcal{S} (\boldsymbol{Z}^\mathcal{S},C^\mathcal{S}|\boldsymbol{X}^\mathcal{S})}  \log p_\mathcal{S} (\boldsymbol{X}^\mathcal{S}|\boldsymbol{Z}^\mathcal{S}) \\
& s.t.\,\, {\rm KL} ( q_\mathcal{S}(\boldsymbol{Z}^\mathcal{S},C^\mathcal{S}|\boldsymbol{X}^\mathcal{S})  || p_\mathcal{S} (\boldsymbol{Z}^\mathcal{S},\boldsymbol{C}^\mathcal{S} ) ) < \delta, 
\end{aligned}
\end{equation}
where $\delta$ specifies the strength of the KL constraint.
Rewriting the optimization problem in Equation~\eqref{equ:opt} under the KKT conditions and neglecting the small number of $\epsilon$, we have:
\begin{equation}
\begin{aligned}
\mathcal{L}_{ELBO}(\boldsymbol{X}^\mathcal{S},\beta) &= \mathbb{E}_{q_\mathcal{S} (\boldsymbol{Z}^\mathcal{S},C^\mathcal{S}|\boldsymbol{X}^\mathcal{S})}   \log p_\mathcal{S} (\boldsymbol{X}^\mathcal{S}|\boldsymbol{Z}^\mathcal{S} )\\
& -\beta\,{\rm KL} ( q_\mathcal{S}(\boldsymbol{Z}^\mathcal{S},C^\mathcal{S}|\boldsymbol{X}^\mathcal{S})  || p_\mathcal{S} (\boldsymbol{Z}^\mathcal{S},\boldsymbol{C}^\mathcal{S} ) ),
\end{aligned}
\end{equation}
where $\beta$ denotes the hyper-parameter between rating modelling and latent distribution matching \cite{betavae}.

\nosection{Loss computing}
The loss of the variational rating reconstruction module can be deduced from ELBO, which has two main parts, i.e., \textit{rating reconstruction} and \textit{distribution matching}. 
For the rating reconstruction part, the reconstruction loss is given as below: 
\begin{equation}
\begin{aligned}
\mathbb{E}_{q_\mathcal{S} (\boldsymbol{Z}^\mathcal{S},C^\mathcal{S}|\boldsymbol{X}^\mathcal{S})}   \log p_\mathcal{S} (\boldsymbol{X}^\mathcal{S}|\boldsymbol{Z}^\mathcal{S})  = -\frac{1}{N} \sum_{i=1}^N \mathcal{F}(\boldsymbol{X}^\mathcal{S}_i,\hat{\boldsymbol{X}}^\mathcal{S}_i),
\end{aligned}
\end{equation}
where $\mathcal{F}(\cdot,\cdot) $ is the cross-entropy loss. 
For the distribution matching part, it can be further substituted through Kullback-Leibler divergence calculation as below:
\begin{equation}
\myfont{
\begin{aligned}
&{\rm KL} ( q_\mathcal{S}(\boldsymbol{Z}^\mathcal{S},C^\mathcal{S}|\boldsymbol{X}^\mathcal{S})  || p_\mathcal{S} (\boldsymbol{Z}^\mathcal{S},\boldsymbol{C}^\mathcal{S} ) ) \\&= -\frac{1}{2N} \sum_{i=1}^N  \sum_{c=1}^K \gamma^\mathcal{S}_{i,c} \sum_{d=1}^D \left( \log (\breve{\sigma}^\mathcal{S}_{c,d})^2 + \left(\frac{\sigma^\mathcal{S}_{i,d}}{\breve{\sigma}^\mathcal{S}_{c,d}}\right)^2 \right) + \frac{1}{N} \sum_{i=1}^N \sum_{c=1}^K \gamma^\mathcal{S}_{i,c} \log \frac{{\pi}^\mathcal{S}_c}{\gamma^\mathcal{S}_{i,c}} \\& -
\frac{1}{2N} \sum_{i=1}^N \sum_{c=1}^K \gamma^\mathcal{S}_{i,c} \sum_{d=1}^D\left( \frac{\mu^\mathcal{S}_{i,d} - \breve{\mu}^\mathcal{S}_{c,d}}{\breve{\mu}^\mathcal{S}_{c,d}}\right)^2  + \frac{1}{2N} \sum_{i=1}^N \sum_{d=1}^D \left(1 + \log \left(\sigma^\mathcal{S}_{i,d} \right)^2 \right),
\end{aligned}
}
\end{equation}
where $\gamma^\mathcal{S}_c$ represents the probability of belonging to the $c$-th cluster and it can be formulated as:
\begin{equation}
\begin{aligned}
\label{equ:cluster}
\gamma^\mathcal{S}_{i,c} = \frac{{\pi}^\mathcal{S}_c \cdot \mathcal{N} (\boldsymbol{Z}^\mathcal{S}_i| \breve{\boldsymbol{\mu}}^\mathcal{S}_c, (\breve{\boldsymbol{\sigma}}^\mathcal{S}_c)^2 )}{\sum_{c'=1}^K {\pi}^\mathcal{S}_{c'} \cdot \mathcal{N} (\boldsymbol{Z}^\mathcal{S}_i| \breve{\boldsymbol{\mu}}^\mathcal{S}_{c'}, (\breve{\boldsymbol{\sigma}}^\mathcal{S}_{c'})^2 )}.
\end{aligned}
\end{equation}
By the detailed process on the source domain through maximizing ELBO w.r.t. the parameters of $\{{\pi}^\mathcal{S},\breve{\boldsymbol{\mu}}^\mathcal{S},\breve{\boldsymbol{\sigma}}^\mathcal{S},E_\mathcal{S},D_\mathcal{S}\}$, the latent user embeddings will be clustered and become much more discriminative. 
Likewise, we can also optimize ELBO on the target domain parameters $\{{\pi}^\mathcal{T},\breve{\boldsymbol{\mu}}^\mathcal{T},\breve{\boldsymbol{\sigma}}^\mathcal{T},E_\mathcal{T},D_\mathcal{T}\}$.
Therefore, the loss of the whole variational rating reconstruction module is given as:
\begin{equation}
\begin{aligned}
&\min L_{VR} = \min [ - \mathcal{L}_{ELBO}(\boldsymbol{X}^\mathcal{S},\beta) - \mathcal{L}_{ELBO}(\boldsymbol{X}^\mathcal{T},\beta) ] \\& = \frac{1}{N} \sum_{i=1}^N \mathcal{F}(\hat{\boldsymbol{X}}^\mathcal{S}_i, \boldsymbol{X}^\mathcal{S}_i) + \beta\, {\rm KL} (q_\mathcal{S} (\boldsymbol{Z}^\mathcal{S},C^\mathcal{S}|\boldsymbol{X}^\mathcal{S}) || p_\mathcal{S}(\boldsymbol{Z}^\mathcal{S},C^\mathcal{S}) ) \\& + \frac{1}{N} \sum_{i=1}^N \mathcal{F}(\hat{\boldsymbol{X}}^\mathcal{T}_i, \boldsymbol{X}^\mathcal{T}_i) + \beta\, {\rm KL} (q_\mathcal{T} (\boldsymbol{Z}^\mathcal{T},C^\mathcal{T}|\boldsymbol{X}^\mathcal{T}) ||     p_\mathcal{T}(\boldsymbol{Z}^\mathcal{T},C^\mathcal{T}) ).
\end{aligned}
\end{equation}

\subsection{Variational Embedding Alignment Module}
As we have motivated before, the embeddings of similar users should be clustered through variational deep embedding to make them more discriminative, as is shown in Fig.~\ref{fig:insight}. 
However, how to align these Mixture-Of-Gaussian distribution across domains remains a tough challenge. 
In order to tackle this issue, we propose \textit{variational embedding alignment} from both local and global perspectives.
%

\nosection{Variational local embedding alignment}
Variational local embedding alignment in \newmodelname~aims to align the corresponding overlapped user embedding distribution across domains. 
Since variational encoder can measure the mean and variance of the input data, we adopt the 2-Wasserstein distance of different Gaussians distributions to calculate the variational local user embedding alignment loss:
\begin{equation}
\begin{aligned}
L_{VA} &= \sum_{i,j=1}^N  \delta(\boldsymbol{U}^\mathcal{S}_i,\boldsymbol{U}^\mathcal{T}_j)
{d}_W ( \mathcal{N}(\boldsymbol{{\mu}}^\mathcal{S}_i,(\boldsymbol{{\sigma}}^\mathcal{S}_i)^2), \mathcal{N}(\boldsymbol{{\mu}}^\mathcal{T}_j,(\boldsymbol{{\sigma}}^\mathcal{T}_j)^2) )\\
& = \sum_{i,j=1}^N  \delta(\boldsymbol{U}^\mathcal{S}_i,\boldsymbol{U}^\mathcal{T}_j)\left( ||\boldsymbol{{\mu}}^\mathcal{S}_i - \boldsymbol{{\mu}}^\mathcal{T}_j||_2^2 + ||\boldsymbol{{\sigma}}^\mathcal{S}_i - \boldsymbol{{\sigma}}^\mathcal{T}_j||_2^2 \right),
\end{aligned}
\end{equation}
where ${d}_W(\cdot,\cdot)$ denotes the 2-Wasserstein distance of different Gaussians distributions that measures the discrepancy across domains, 
$\delta(\cdot)$ indicates whether a user occurs simultaneously in both domains under the POCDR setting, 
$\delta(\boldsymbol{U}^\mathcal{S}_i,\boldsymbol{U}^\mathcal{T}_j)=1$ means the $i$-th user in the source domain has the same identification with the $j$-th user in the target domain, otherwise $\delta(\boldsymbol{U}^\mathcal{S}_i,\boldsymbol{U}^\mathcal{T}_j) = 0$.
The distribution estimation can further help the model to better capture the uncertainties of the learned embeddings \cite{bivae}.

\nosection{Variational global embedding alignment}
\newmodelname~should also align the global distributions of both the overlapped and non-overlapped users across domains.
Similar intra-domain users can be clustered through variational local embedding alignment, however, there is still a distribution discrepancy between the source and target domains.
Thus, how to match the whole user distributions across domains without explicit label information remains the key issue.

In order to obtain the alignment results effectively, we propose the \textit{Gromov-Wasserstein Distribution Co-Clustering Optimal Transport} (GDOT) approach.
Unlike conventional optimal transport \cite{santambrogio2015optimal,villani2008optimal} that only aligns user embeddings, our proposed GDOT for domain adaptation performs user embedding distribution alignment in the source and target domains. 
The optimization of optimal transport is based on the Kantorovich problem \cite{angenent2003minimizing}, which seeks a general coupling $\psi \in \mathcal{X}(\mathcal{S},\mathcal{T})$ between $\mathcal{S}$ and $\mathcal{T}$:
\begin{equation}
\begin{aligned}
\psi^* = \mathop{\arg\min}_{\psi \in \mathcal{X}(\mathcal{S},\mathcal{T})}  \iint_{\mathcal{S} \times \mathcal{T}} \mathcal{M}({\mathbb{P}}^\mathcal{S}_{i,i'},{\mathbb{P}}^\mathcal{T}_{j,j'}) \,d\psi({\mathbb{P}}^\mathcal{S}_i,{\mathbb{P}}^\mathcal{T}_j)d\psi({\mathbb{P}}^\mathcal{S}_{i'},{\mathbb{P}}^\mathcal{T}_{j'}),
\end{aligned}
\end{equation}
where $\mathcal{X}(\mathcal{S}, \mathcal{T})$ denotes the latent embedding probability distribution between $\mathcal{S}$ and $\mathcal{T}$. 
The cost function matrix $\mathcal{M}(\mathbb{P}^\mathcal{S}_{i,i'},\mathbb{P}^\mathcal{T}_{j,j'})$ measures the difference between each pair of distributions.
To facilitate calculation, we formulate the discrete optimal transport formulation as $\psi^* = \mathop{\arg\min}_{\boldsymbol{\psi}\boldsymbol{1}_K = \boldsymbol{\pi}^\mathcal{S}, \boldsymbol{\psi}^T\boldsymbol{1}_K = \boldsymbol{\pi}^\mathcal{T} } \langle \mathcal{{M}} \otimes \boldsymbol{\psi}, \boldsymbol{\psi} \rangle_{F} - \epsilon H(\boldsymbol{\psi})$, where $\boldsymbol{\psi}^* \in \mathbb{R} ^ {K \times K}$ is the ideal coupling matrix between the source and target domains, representing the joint probability of the source probability $\mathbb{P}^\mathcal{S}$ and the target probability $\mathbb{P}^\mathcal{T}$.
The matrix $\mathcal{{M}} \in \mathbb{R}^{K \times K \times K \times K}$ denotes a pairwise Gaussian distribution distance among the source probability $\mathbb{P}^\mathcal{S}_i,\mathbb{P}^\mathcal{S}_{i'}$ and target probability $\mathbb{P}^\mathcal{T}_j,\mathbb{P}^\mathcal{T}_{j'}$ as:
\begin{equation}
\begin{aligned}
\label{equ:get_m}
\mathcal{{M}}[i][j][i'][j'] = ||{d}_W(\mathbb{P}^\mathcal{S}_i,\mathbb{P}^\mathcal{S}_{i'}) - {d}_W(\mathbb{P}^\mathcal{T}_j,\mathbb{P}^\mathcal{T}_{j'})||_2^2.
\end{aligned}
\end{equation}
Meanwhile $\otimes$ denotes the tensor matrix multiplication, which means $[\mathcal{{M}} \otimes \boldsymbol{\psi}]_{ij} = (\sum_{i',j'} \mathcal{{M}}_{i,j,i',j'} \boldsymbol{\psi}_{i',j'})_{i,j}$.
The second term $H(\boldsymbol{\psi}) = -\sum_{i=1}^K \sum_{j=1}^K \psi_{ij}(\log(\psi_{ij}) - 1)$ is a regularization term and $\epsilon$ is a balance hyper parameter. 
The matching matrix follows the constraints that $\boldsymbol{\psi}\boldsymbol{1}_K = \boldsymbol{\pi}^\mathcal{S}, \boldsymbol{\psi}^T\boldsymbol{1}_K = \boldsymbol{\pi}^\mathcal{T}$ with $K$ denoting the number of clusters.
We apply the Sinkhorn algorithm \cite{sinkhorn,icml2020} for solving the optimization problem by adopting the Lagrangian multiplier to minimize the objective function $\ell$ as below:
\begin{equation}
\begin{aligned}
\label{equ:ell}
\ell= \langle \mathcal{{M}} \otimes \boldsymbol{\psi}, \boldsymbol{\psi} \rangle - \epsilon H(\boldsymbol{\psi}) - \boldsymbol{f}(\boldsymbol{\psi}\boldsymbol{1}_K - \boldsymbol{\pi}^\mathcal{S}) - \boldsymbol{g}(\boldsymbol{\psi}^T\boldsymbol{1}_K - \boldsymbol{\pi}^\mathcal{T}),
\end{aligned}
\end{equation}
where $\boldsymbol{f}$ and $\boldsymbol{g}$ are both Lagrangian multipliers. 
We first randomly generate a matrix $\boldsymbol{\psi}$ that meets the boundary constraints and then define the variable $\mathcal{G} = \mathcal{M} \otimes \mathcal{\psi}$.
Taking the differentiation of Equation~\eqref{equ:ell} on variable $\boldsymbol{\psi}$, we can obtain:
\begin{equation}
\begin{aligned}
\label{equ:pi}
\psi_{ij} = \exp \left( \frac{f_i}{\epsilon}\right)\exp \left( -\frac{\mathcal{G}_{ij}}{\epsilon}\right)\exp \left( \frac{g_j}{\epsilon}\right).
\end{aligned}
\end{equation}
Taking Equation~\eqref{equ:pi} back to the original constraints $\boldsymbol{\psi}\boldsymbol{1}_K = \boldsymbol{\pi}^\mathcal{S}$ and $ \boldsymbol{\psi}^T\boldsymbol{1}_K = \boldsymbol{\pi}^\mathcal{T}$, we can obtain that:
\begin{equation}
\begin{aligned}
\label{equ:sinkhorn}
\boldsymbol{u} \odot (\boldsymbol{Hv}) = \boldsymbol{\pi}^\mathcal{S},\,\,\boldsymbol{v} \odot (\boldsymbol{H}^T\boldsymbol{u}) = \boldsymbol{\pi}^\mathcal{T}.
\end{aligned}
\end{equation}
Since we define $\boldsymbol{u} = {\rm diag}\left( \frac{\boldsymbol{f}}{\epsilon}\right)$, $\boldsymbol{H} = \exp \left( -\frac{\mathcal{{G}}}{\epsilon}\right)$, and $\boldsymbol{v} = {\rm diag}\left( \frac{\boldsymbol{g}}{\epsilon}\right)$.
Through iteratively solving Equation~\eqref{equ:sinkhorn} on the variables $\boldsymbol{u}$ and $\boldsymbol{v}$, we can finally obtain the coupling matrix $\boldsymbol{\pi}^*$ on Equation~\eqref{equ:pi}.
We show the iteration optimization scheme of GDOT in Algorithm~1. 
In summary, variational embedding alignment can obtain the optimal transport solution as $\boldsymbol{\psi}^*$ for each corresponding sub-distribution, as shown with dot lines in Fig.~\ref{fig:insight}(b). 

\begin{algorithm}[htbp]
\begin{algorithmic}[1]
{
    \STATE {
    \textbf{Input:} $K$: Number of clusters; $\mathcal{N}(\boldsymbol{\breve{\mu}}^\mathcal{S}_c,(\boldsymbol{\breve{\sigma}}^\mathcal{S}_c)^2)$: The source user latent distribution for the $c$-th cluster group; $\mathcal{N}(\boldsymbol{\breve{\mu}}^\mathcal{T}_c,(\boldsymbol{\breve{\sigma}}^\mathcal{T}_c)^2)$: The target user latent distribution for the $c$-th cluster group; $\epsilon$: Hyper parameter to balance the matching loss and the entropy maximization constraint.
    }
    \STATE {
    Initialize the coupling matrix $\boldsymbol{\psi}$.
    }
    \FOR{$i=1$ to $T$}
    \STATE {Calculate the distance between different Normal distribution in source and target domains as matrix ${\mathcal{M}} \in \mathbb{R}^{K \times K \times K \times K}$ through Equation \eqref{equ:get_m}.}
    \STATE {Calculate matrix $\mathcal{G} = {\mathcal{M}} \otimes \boldsymbol{\psi}$ and $\boldsymbol{H} = \exp \left( -\frac{\mathcal{{G}}}{\epsilon}\right)$. }
    \STATE {Randomly initialize vectors $\boldsymbol{u}\in \mathbb{R}^K$ and $\boldsymbol{v}\in \mathbb{R}^K$.}
        \FOR{$j=1$ to $t$}
            \STATE {Update $\boldsymbol{u}  \leftarrow \boldsymbol{\pi}^{\mathcal{S}} \oslash (\boldsymbol{H}\boldsymbol{v})$. }
            \STATE {Update $\boldsymbol{v} \leftarrow \boldsymbol{\pi}^{\mathcal{T}} \oslash (\boldsymbol{H}^{\top}\boldsymbol{u})$. }
        \ENDFOR
        \STATE{$\boldsymbol{\psi} = {\rm diag}(\boldsymbol{u}) \boldsymbol{H} {\rm diag}(\boldsymbol{v})$. }
    \ENDFOR
    \STATE{\textbf{Return}: $\boldsymbol{\psi}^* = \boldsymbol{\psi}$. }
}
\end{algorithmic}
\caption{The optimization scheme of GDOT.}
\end{algorithm}

After that, the variational global embedding alignment loss is given as:
\begin{equation}
\begin{aligned}
L_{VG} = \sum_{i=1}^K \sum_{j=1}^K \psi^*_{ij} d_W \left (\left(\mathcal{N}\left(\breve{\boldsymbol{\mu}}^\mathcal{S}_i,(\breve{\boldsymbol{\sigma}}^\mathcal{S}_i)^2\right),\mathcal{N}\left(\breve{\boldsymbol{\mu}}^\mathcal{T}_j,(\breve{\boldsymbol{\sigma}}^\mathcal{T}_j)^2\right) \right) \right). 
\end{aligned}
\end{equation}
The bigger $\psi^*_{ij}$ for the matching constraint, the closer between the distributions $\mathcal{N}(\breve{\boldsymbol{\mu}}^\mathcal{S}_i,(\breve{\boldsymbol{\sigma}}^\mathcal{S}_i)^2)$ and $\mathcal{N}(\breve{\boldsymbol{\mu}}^\mathcal{T}_j,(\breve{\boldsymbol{\sigma}}^\mathcal{T}_j)^2)$.

\subsection{Putting Together}
The total loss of \newmodelname~could be obtained by combining the losses of the variational rating reconstruction module and the variational embedding alignment module.
That is, the losses of \newmodelname~is given as:
\begin{equation}
\begin{aligned}
\label{equ:total}
L_{\newmodelname} = L_{VR} + \lambda_{VL} L_{VA} + \lambda_{VG} L_{VG}, 
\end{aligned}
\end{equation}
where $\lambda_{VL}$ and $\lambda_{VG}$ are hyper-parameters to balance different type of losses.
By doing this, users with similar rating behaviors will have close embedding distance in single domain through the variational inference, and the user embeddings across domains will become domain-invariant through GDOT which has been shown in Fig.~\ref{fig:insight}(c). 

\section{Empirical Study}


In this section, we conduct experiments on several real-world datasets to answer the following questions: 
(1) \textbf{RQ1}: How does our approach perform compared with the state-of-the-art single-domain or cross-domain recommendation methods?
(2) \textbf{RQ2}: How do the local and global embedding alignments contribute to performance improvement? Can proposed GDOT better align the global distribution?
(3) \textbf{RQ3}: How does the performance of \newmodelname~vary with different values of the hyper-parameters?

\subsection{Datasets and Tasks} 
We conduct extensive experiments on two popularly used real-world datasets, i.e., \textit{Douban} and \textit{Amazon}, for evaluating our models on POCDR tasks \cite{dtcdr,gadtcdr,amazon}.
First, the \textbf{Douban} dataset \cite{dtcdr,gadtcdr} has two domains, i.e., Book and Movie, with user-item ratings. 
Second, the \textbf{Amazon} dataset \cite{catn,amazon} has four domains, i.e., Movies and TV (Movie), Books (Book), CDs and Vinyl (Music), and Clothes (Clothes).
The detailed statistics of these datasets after pre-process will be shown in Table 1.
For both datasets, we binarize the ratings to 0 and 1.
Specifically, we take the ratings higher or equal to 4 as positive and others as 0. 
We also filter the users and items with less than 5 interactions, following existing research \cite{darec,gadtcdr}.
We provide four main tasks to evaluate our model, i.e., \textbf{Amazon Movie} $\&$ \textbf{Amazon Music}, \textbf{Amazon Movie} $\&$ \textbf{Amazon Book}, \textbf{Amazon Movie} $\&$ \textbf{Amazon Clothes}, and \textbf{Douban Movie} $\&$ \textbf{Douban Book}.
It is noticeable that some tasks are rather simple because the source and target domains are similar, e.g., \textbf{Amazon Movie} $\&$ \textbf{Amazon Book}, while some tasks are difficult since these two domains are different, e.g., \textbf{Amazon Movie} $\&$ \textbf{Amazon Clothes} \cite{tdar}.

\begin{table}[t]
\small
  \centering
  \caption{Statistics on Douban and Amazon datasets.}
    \begin{tabular}{cccccc}
    \hline
    
    \multicolumn{2}{c}{ \textbf{Datasets} } & \textbf{Users} & \textbf{Items} &  \textbf{Ratings} & \textbf{Density}\\
    
    \hline
    
    \multirow{2}{*}{\textbf{Amazon}} & Movie & \multirow{2}{*}{15,914} & 17,794     & 416,228    & 0.14\% \\
          & Music &       & 20,058    & 280,398     & 0.09\% \\

\hline

    \multirow{2}{*}{\textbf{Amazon}} & Movie & \multirow{2}{*}{29,476} &  24,091     &  591,258     & 0.08\% \\
         & Book  &       &  41,884     &  579,131     & 0.05\% \\
    
    \hline
    
    \multirow{2}{*}{\textbf{Amazon}} & Movie & \multirow{2}{*}{13,267} &  20,433     &  216,868     & 0.08\%  \\
         & Clothes  &       & 21,054      &  195,550     & 0.07\% \\
    
    \hline
    
    \multirow{2}{*}{\textbf{Douban}} & Movie & \multirow{2}{*}{2,106} & 9,551      & 871,280      &  4.33\%\\
          & Book  &       & 6,766      &  90,164     & 0.63\% \\
\hline

    \end{tabular}%
  \label{tab:datasetss}%
\end{table}%
 \vspace{-0.3cm}

\subsection{Experimental Settings}

%
We randomly divide the observed source and target data into training, validation, and test sets with a ratio of 8:1:1.
Meanwhile, we vary the overlapped user ratio $\mathcal{K}_u$ in $\{30\%,60\%,90\%\}$.
Different user overlapped ratio represents different situations, e.g., $\mathcal{K}_u = 30\%$ represents only few users are overlapped while $\mathcal{K}_u = 90\%$ means most of users are overlapped \cite{kerkt}.
%
%
For all the experiments, we perform five random experiments and report the average results.
We choose Adam \cite{Adam} as optimizer, and adopt Hit Rate$@k$ (HR$@k$) and NDCG$@k$ \cite{kat} as the ranking evaluation metrics with $k = 5$.

\begin{table*}[htbp]
\renewcommand\arraystretch{1}
\small
  \centering
  \caption{Experimental results on Amazon and Douban datasets.}
    \begin{tabular}{cccccccccccccc}
    \bottomrule

    \multicolumn{2}{c}{} & 
    \multicolumn{6}{c}{(Amazon) Movie \& Music} & 
    \multicolumn{6}{c}{(Amazon) Movie \& Clothes} \\
    
    \cmidrule(lr){3-8}
    \cmidrule(lr){9-14}
    
    \multicolumn{2}{c}{} & 
    \multicolumn{2}{c}{$\mathcal{K}_u = 30\%$} & \multicolumn{2}{c}{$\mathcal{K}_u = 60\%$} & \multicolumn{2}{c}{$\mathcal{K}_u = 90\%$} &
    \multicolumn{2}{c}{$\mathcal{K}_u = 30\%$} & \multicolumn{2}{c}{$\mathcal{K}_u = 60\%$} & \multicolumn{2}{c}{$\mathcal{K}_u = 90\%$} 
    \\
    
    \multicolumn{2}{c}{} & \multicolumn{1}{c}{Movie} & \multicolumn{1}{c}{Music} & \multicolumn{1}{c}{Movie} & \multicolumn{1}{c}{Music} &  \multicolumn{1}{c}{Movie} & \multicolumn{1}{c}{Music} & \multicolumn{1}{c}{Movie} & \multicolumn{1}{c}{Clothes} & \multicolumn{1}{c}{Movie} & \multicolumn{1}{c}{Clothes} &
    \multicolumn{1}{c}{Movie} & \multicolumn{1}{c}{Clothes} \\
    
    \hline
    
    \multirow{2}{*}{KerKT} & HR    & 0.3463 & 0.3505 & 0.3750     & 0.3936     & 0.4228     & 0.4409     & 0.2155     & 0.0876     & 0.2614     & 0.1165 & 0.3155 & 0.1648     \\
    
          & NDCG  & 0.2504  & 0.2572 & 0.2858     & 0.2994     & 0.3253     & 0.3352     & 0.1423     & 0.0983     & 0.1867     & 0.1285    & 0.2425  & 0.1676 \\
    
    \hline

    \multirow{2}{*}{NeuMF} & HR    & 0.3661 & 0.3722 & 0.3987     & 0.4128     & 0.4467     & 0.4661     & 0.2281    & 0.0990     & 0.2697     & 0.1382 & 0.3307 & 0.1820    \\
    
          & NDCG  & 0.2602 & 0.2686 & 0.2944     & 0.3109     & 0.3319     & 0.3488     & 0.1606     & 0.1081     & 0.1980     & 0.1361 & 0.2570 & 0.1774   \\
          
    \hline

    \multirow{2}{*}{MultVAE} & HR    & 0.3807 & 0.3861 & 0.4019     & 0.4247     & 0.4526     & 0.4768     & 0.2373    & 0.1047     & 0.2802     & 0.1435 & 0.3451 & 0.1890    \\
    
          & NDCG  & 0.2654 & 0.2758 & 0.2995     & 0.3346     & 0.3372     & 0.3649     & 0.1686     & 0.1161     & 0.2084     & 0.1490 & 0.2627 & 0.1868   \\
          
    \hline

    \multirow{2}{*}{BiVAE} & HR    & 0.3880 & 0.3956 & 0.4104     & 0.4392     & 0.4621     & 0.4833     & 0.2436    & 0.1152     & 0.2934     & 0.1627 & 0.3553 & 0.1991    \\
    
          & NDCG  & 0.2754 & 0.2836 & 0.3109     & 0.3365     & 0.3417     & 0.3758     & 0.1817     & 0.1266     & 0.2218     & 0.1615 & 0.2754 & 0.1989   \\
    
    \hline

    \multirow{2}{*}{DARec} & HR    & 0.3831 & 0.3898 & 0.4055     & 0.4311     & 0.4560     & 0.4779     & 0.2414    & 0.1084     & 0.2878     & 0.1509 & 0.3505 & 0.1940    \\
    
          & NDCG  & 0.2699 & 0.2740 & 0.3038     & 0.3386     & 0.3423     & 0.3689     & 0.1754     & 0.1216     & 0.2162     & 0.1565 & 0.2697 & 0.1973   \\
         
    \hline

    \multirow{2}{*}{ETL} & HR    & 0.3893 & 0.4002 & 0.4160     & 0.4434     & 0.4629     & 0.4858     & 0.2508    & 0.1160     & 0.3019     & 0.1683 & 0.3632 & 0.2057    \\
    
          & NDCG  & 0.2751 & 0.2898 & 0.3145     & 0.3393     & 0.3455     & 0.3781     & 0.1872     & 0.1294     & 0.2295     & 0.1720 & 0.2818 & 0.2036   \\
          
              \hline

    \multirow{2}{*}{DML} & HR    & 0.3941 & 0.4064 & 0.4182     & 0.4510     & 0.4676     & 0.4885     & 0.2552    & 0.1201     & 0.3093     & 0.1714 & 0.3674 & 0.2065    \\
    
          & NDCG  & 0.2763 & 0.2954 & 0.3156     & 0.3425     & 0.3487     & 0.3778     & 0.1926     & 0.1335     & 0.2349     & 0.1672 & 0.2861 & 0.2050   \\
     
     \hline 
     
    \multirow{2}{*}{\newmodelname} & HR    & \textbf{0.4404} & \textbf{0.4457} & \textbf{0.4602}     & \textbf{0.4865}     & \textbf{0.4956}     & \textbf{0.5148}     &   \textbf{0.2867}  & \textbf{0.1485}     & \textbf{0.3342}     & \textbf{0.1944} & \textbf{0.3884} & \textbf{0.2303}    \\
    
          & NDCG  & \textbf{0.3021} & \textbf{0.3207} & \textbf{0.3322}     & \textbf{0.3645}     & \textbf{0.3659}     & \textbf{0.3966}     & \textbf{0.2231}     & \textbf{0.1589}     & \textbf{0.2538}     & \textbf{0.1869} & \textbf{0.2973} & \textbf{0.2287}   \\          
     \bottomrule

    \multicolumn{2}{c}{} & 
    \multicolumn{6}{c}{(Amazon) Movie \& Book} & 
    \multicolumn{6}{c}{(Douban) Movie \& Book} \\
    
    \cmidrule(lr){3-8}
    \cmidrule(lr){9-14}
    
    \multicolumn{2}{c}{} & 
    \multicolumn{2}{c}{$\mathcal{K}_u = 30\%$} & \multicolumn{2}{c}{$\mathcal{K}_u = 60\%$} & \multicolumn{2}{c}{$\mathcal{K}_u = 90\%$} &
    \multicolumn{2}{c}{$\mathcal{K}_u = 30\%$} & \multicolumn{2}{c}{$\mathcal{K}_u = 60\%$} & \multicolumn{2}{c}{$\mathcal{K}_u = 90\%$} 
    \\
    
    \multicolumn{2}{c}{} & \multicolumn{1}{c}{Movie} & \multicolumn{1}{c}{Book} & \multicolumn{1}{c}{Movie} & \multicolumn{1}{c}{Book} &  \multicolumn{1}{c}{Movie} & \multicolumn{1}{c}{Book} & \multicolumn{1}{c}{Movie} & \multicolumn{1}{c}{Book} & \multicolumn{1}{c}{Movie} & \multicolumn{1}{c}{Book} &
    \multicolumn{1}{c}{Movie} & \multicolumn{1}{c}{Book} \\
    
    \hline
    
    \multirow{2}{*}{KerKT} & HR    & 0.3507 & 0.3534 & 0.3903     & 0.3958     & 0.4448     & 0.4421     & 0.4518     & 0.3009     & 0.5094     & 0.3353 & 0.5496 & 0.3657     \\
    
          & NDCG  & 0.2505  & 0.2559 & 0.2947     & 0.3013     & 0.3266     & 0.3358     & 0.2529     & 0.1392     & 0.3171     & 0.1705    & 0.3324  & 0.2006 \\
    
    \hline

    \multirow{2}{*}{NeuMF} & HR    & 0.3722 & 0.3774 & 0.4155     & 0.4183     & 0.4607     & 0.4656     & 0.4610    & 0.3253     & 0.5247     & 0.3526 & 0.5659 & 0.3881    \\
    
          & NDCG  & 0.2638 & 0.2672 & 0.3021     & 0.3094     & 0.3356     & 0.3457     & 0.2656     & 0.1450     & 0.3291     & 0.1763 & 0.3523 & 0.2082   \\
    
    \hline

    \multirow{2}{*}{MultVAE} & HR    & 0.3878 & 0.3940 & 0.4239     & 0.4217     & 0.4682     & 0.4751     & 0.4793    & 0.3503     & 0.5312     & 0.3864 & 0.5780 & 0.4044    \\
    
          & NDCG  & 0.2763 & 0.2728 & 0.3124     & 0.3108     & 0.3440     & 0.3539     & 0.2745     & 0.1627     & 0.3441     & 0.1952 & 0.3724 & 0.2163   \\
          
    \hline

    \multirow{2}{*}{BiVAE} & HR    & 0.3983 & 0.3992 & 0.4426     & 0.4501     & 0.4837     & 0.4851     & 0.4921    & 0.3619     & 0.5586     & 0.4208 & 0.6014 & 0.4393    \\
    
          & NDCG  & 0.2852 & 0.2837 & 0.3289     & 0.3324     & 0.3520     & 0.3613     & 0.2945     & 0.1728     & 0.3664     & 0.2326 & 0.3935 & 0.2437   \\
    
    \hline

    \multirow{2}{*}{DARec} & HR    & 0.3947 & 0.3918 & 0.4364     & 0.4350     & 0.4748     & 0.4795     & 0.4859    & 0.3538     & 0.5445     & 0.3918 & 0.5912 & 0.4251    \\
    
          & NDCG  & 0.2816 & 0.2783 & 0.3230     & 0.3297     & 0.3484     & 0.3572     & 0.2797     & 0.1668     & 0.3563     & 0.2035 & 0.3819 & 0.2286   \\
         
    \hline

    \multirow{2}{*}{ETL} & HR    & 0.4022 & 0.4075 & 0.4269     & 0.4538     & 0.4883     & 0.4905     & 0.5044    & 0.3706     & 0.4269     & 0.4538 & 0.6175 & 0.4503    \\
    
          & NDCG  & 0.2959 & 0.2893 & 0.3341     & 0.3386     & 0.3590     & 0.3724     & 0.3088     & 0.1817     & 0.3715     & 0.2314 & 0.3992 & 0.2496   \\
          
              \hline

    \multirow{2}{*}{DML} & HR    & 0.4070
    & 0.4013 & 0.4522 & 0.4557     & 0.4932     & 0.4941     & 0.5187    & 0.3725     & 0.5805     & 0.4298 & 0.6284 & 0.4618    \\
    
          & NDCG  & 0.3073 & 0.2934 & 0.3389     & 0.3426     & 0.3642     & 0.3716     & 0.3176     & 0.1884     & 0.3785     & 0.2332 & 0.4025 & 0.2551   \\
          
          \hline
          
    \multirow{2}{*}{\newmodelname} & HR    & \textbf{0.4481} & \textbf{0.4342} & \textbf{0.4813}     & \textbf{0.4824}     & \textbf{0.5145}     & \textbf{0.5058}     & \textbf{0.5662}    & \textbf{0.4023}     & \textbf{0.6009}     & \textbf{0.4345} & \textbf{0.6435} & \textbf{0.4709}   \\
    
          & NDCG  & \textbf{0.3338} & \textbf{0.3206} & \textbf{0.3581}     & \textbf{0.3612}    & \textbf{0.3889}     & \textbf{0.3857}     & \textbf{0.3519}     & \textbf{0.2150}     & \textbf{0.3903}     & \textbf{0.2421} & \textbf{0.4207} & \textbf{0.2644}  \\   
          
     \bottomrule
     
    \vspace{-0.3cm}
    \end{tabular}%
  \label{tab:amazon1}%
\end{table*}%

\subsection{Baseline}
We compare our proposed \textbf{\newmodelname~}with the following state-of-the-art cross domain recommendation models.
(1) \textbf{KerKT} \cite{kerkt} Kernel-induced knowledge transfer for overlapping entities (KerKT) which is the first attempt on POCDR problem. It uses a shallow model with multiple steps to model and align user representations.
%
%
(2) \textbf{NeuMF} \cite{neumf} Neural Collaborative Filtering for Personalized Ranking (NeuMF) replaces the traditional inner product with a neural network to learn an arbitrary function on the single domain accordingly. 
(3) \textbf{MultVAE} \cite{multvae} Variational Autoencoders for Collaborative Filtering (MultVAE) extends variational autoencoders for recommendation.
(4) \textbf{BiVAE} \cite{bivae} Bilateral Variational Autoencoder for Collaborative Filtering (BiVAE) further adopts variational autoencoder on both user and item to model the rating interactions.
(5) \textbf{DARec} \cite{darec} Deep Domain Adaptation for Cross-Domain Recommendation via Transferring Rating Patterns (DARec) adopts adversarial training strategy to extract and transfer knowledge patterns for shared users across domains.
(6) \textbf{ETL} \cite{etl} Equivalent Transformation Learner (ETL) of user preference in CDR adopts equivalent transformation module to capture both the overlapped and domain-specific properties.
(7) \textbf{DML} \cite{dml} Dual Metric Learning for Cross-Domain Recommendations (DML) is the recent state-of-the-art CDR model which transfers information across domains based on dual modelling.

\subsection{Implementation Details}

We set batch size $N=256$ for both the source and target domains.
The latent embedding dimension is set to $D = 128$.
We assume both source and target users share the same embedding dimension according to previous research \cite{unit,etl}.
For the \textit{variational rating reconstruction module}, we adopt the heuristic strategy for selecting the disentangle factor $\beta$ by starting from $\beta = 0$ and gradually increasing to $\beta = 0.2$ through KL annealing \cite{multvae}. 
Empirically, we set the cluster number as $K = 30$.
Meanwhile, we set hyper-parameters $\epsilon = 0.1$ for Gromov-Wasserstein Distribution Co-Clustering Optimal Transport.
For \newmodelname~model, we set the balance hyper-parameters as $\lambda_{VL} = 0.7$ and $\lambda_{VG} = 1.0$.



\begin{table*}
\small
  \centering
  \caption{Ablation results on Amazon datasets. }
  \label{tab:ablation}
    \begin{tabular}{cccccccccccccc}
    \bottomrule

    \multicolumn{2}{c}{} & 
    \multicolumn{4}{c}{(Amazon) Movie \& Music} & 
    \multicolumn{4}{c}{(Amazon) Movie \& Book} & 
    \multicolumn{4}{c}{(Amazon) Movie \& Clothes} \\
    
    \cmidrule(lr){3-6}
    \cmidrule(lr){7-10}
    \cmidrule(lr){11-14}
    
    \multicolumn{2}{c}{} & 
    \multicolumn{2}{c}{$\mathcal{K}_u = 30\%$} & \multicolumn{2}{c}{$\mathcal{K}_u = 90\%$} & \multicolumn{2}{c}{$\mathcal{K}_u = 30\%$} &
    \multicolumn{2}{c}{$\mathcal{K}_u = 90\%$} & \multicolumn{2}{c}{$\mathcal{K}_u = 30\%$} & \multicolumn{2}{c}{$\mathcal{K}_u = 90\%$} 
    \\
    
    \multicolumn{2}{c}{} & \multicolumn{1}{c}{Movie} & \multicolumn{1}{c}{Music} & \multicolumn{1}{c}{Movie} & \multicolumn{1}{c}{Music} &  \multicolumn{1}{c}{Movie} & \multicolumn{1}{c}{Book} & \multicolumn{1}{c}{Movie} & \multicolumn{1}{c}{Book} & \multicolumn{1}{c}{Movie} & \multicolumn{1}{c}{Clothes} &
    \multicolumn{1}{c}{Movie} & \multicolumn{1}{c}{Clothes} \\
    
    \hline
    
    \multirow{2}{*}{\newmodelname-Base} & HR    & 0.3901 & 0.3994 & 0.4597     & 0.4822     & 0.3895     & 0.4002     & 0.4816     & 0.4789     & 0.2467     & 0.1169 & 0.3613 & 0.1944     \\
    
          & NDCG  & 0.2710  & 0.2805 & 0.3368     & 0.3723     & 0.2816     & 0.2862     & 0.3485     & 0.3591     & 0.1826     & 0.1245    & 0.2703  & 0.1938 \\
     
     \hline 
     
    \multirow{2}{*}{\newmodelname-Local} & HR    & 0.4217 & 0.4238 & 0.4854     & 0.5034     & 0.4295     & 0.4305     & 0.5037     & 0.4965     & 0.2718     & 0.1373 & 0.3779 & 0.2194     \\
    
          & NDCG  & 0.2889  & 0.3076 & 0.3561     & 0.3850     & 0.3219     & 0.3158     & 0.3787     & 0.3793     & 0.2084     & 0.1497    & 0.2864  & 0.2198 \\
          \hline

     \multirow{2}{*}{\newmodelname-Global} & HR    & 0.4330 & 0.4315 & 0.4913     & 0.5082     & 0.4386     & 0.4277     & 0.5092     & 0.5017     & 0.2786     & 0.1402 & 0.3826 & 0.2259     \\
    
          & NDCG  & 0.2936  & 0.3134 & 0.3614     & 0.3901     & 0.3274     & 0.3129     & 0.3816     & 0.3822     & 0.2145     & 0.1510    & 0.2916  & 0.2231 \\
     
     \hline 
     
    \multirow{2}{*}{\newmodelname} & HR    & \textbf{0.4404} & \textbf{0.4456} & \textbf{0.4956}     & \textbf{0.5147}     & \textbf{0.4481}     & \textbf{0.4342}     &  \textbf{0.5145}   & \textbf{0.5058}     & \textbf{0.2867}     & \textbf{0.1485} & \textbf{0.3884} & \textbf{0.2303}    \\
    
          & NDCG  & \textbf{0.3021} & \textbf{0.3207} & \textbf{0.3659}     & \textbf{0.3966}     & \textbf{0.3338}     & \textbf{0.3206}     & \textbf{0.3889}    & \textbf{0.3857}     & \textbf{0.2231}     & \textbf{0.1589} & \textbf{0.2973} & \textbf{0.2287}   \\          
     \bottomrule

    \end{tabular}%
  \label{tab:amazon2}%
\end{table*}%

\begin{figure*} 
    \centering
    
    \subfigure[\newmodelname-Base]{
    \begin{minipage}[t]{0.3\linewidth} 
    \includegraphics[width=5.4cm]{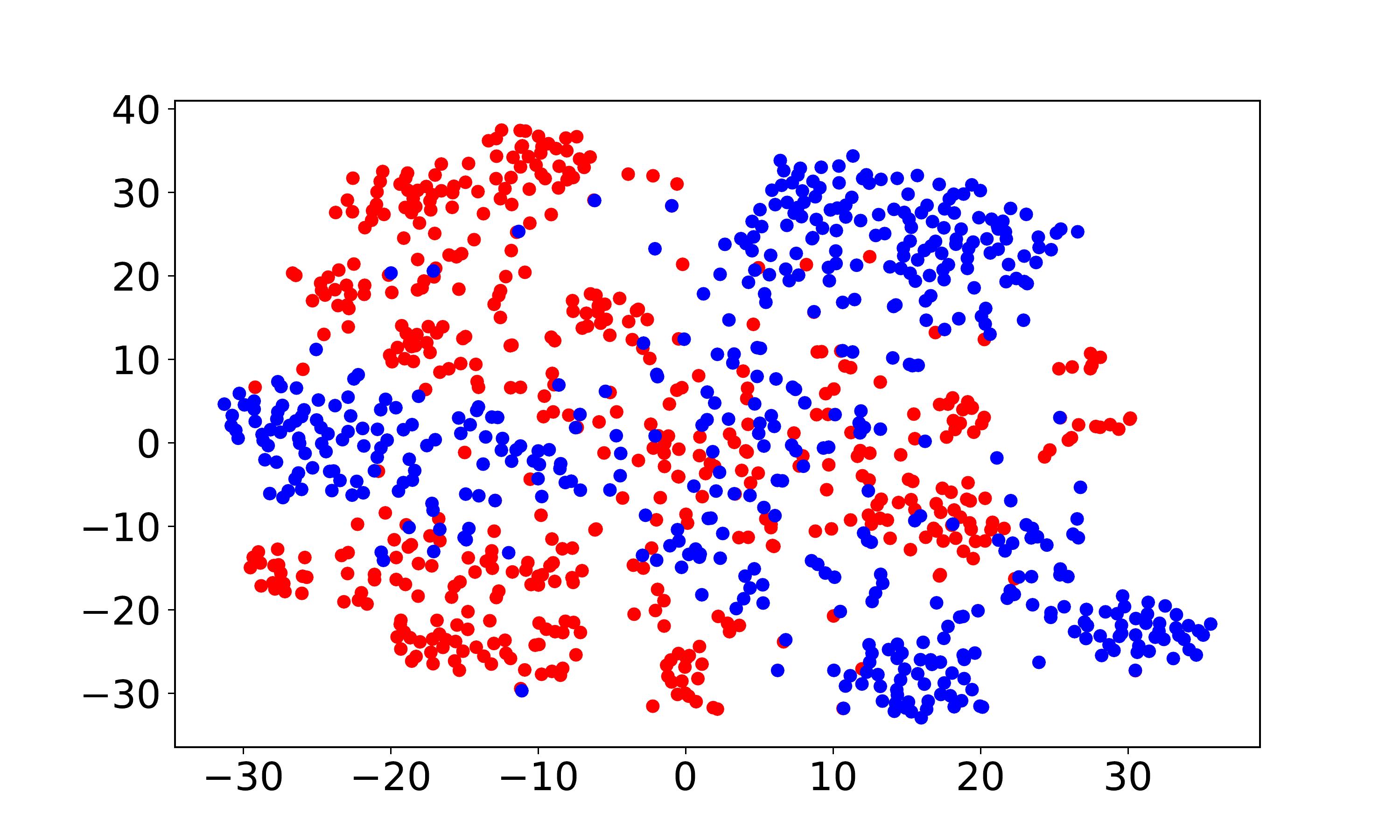}\\
    \includegraphics[width=5.4cm]{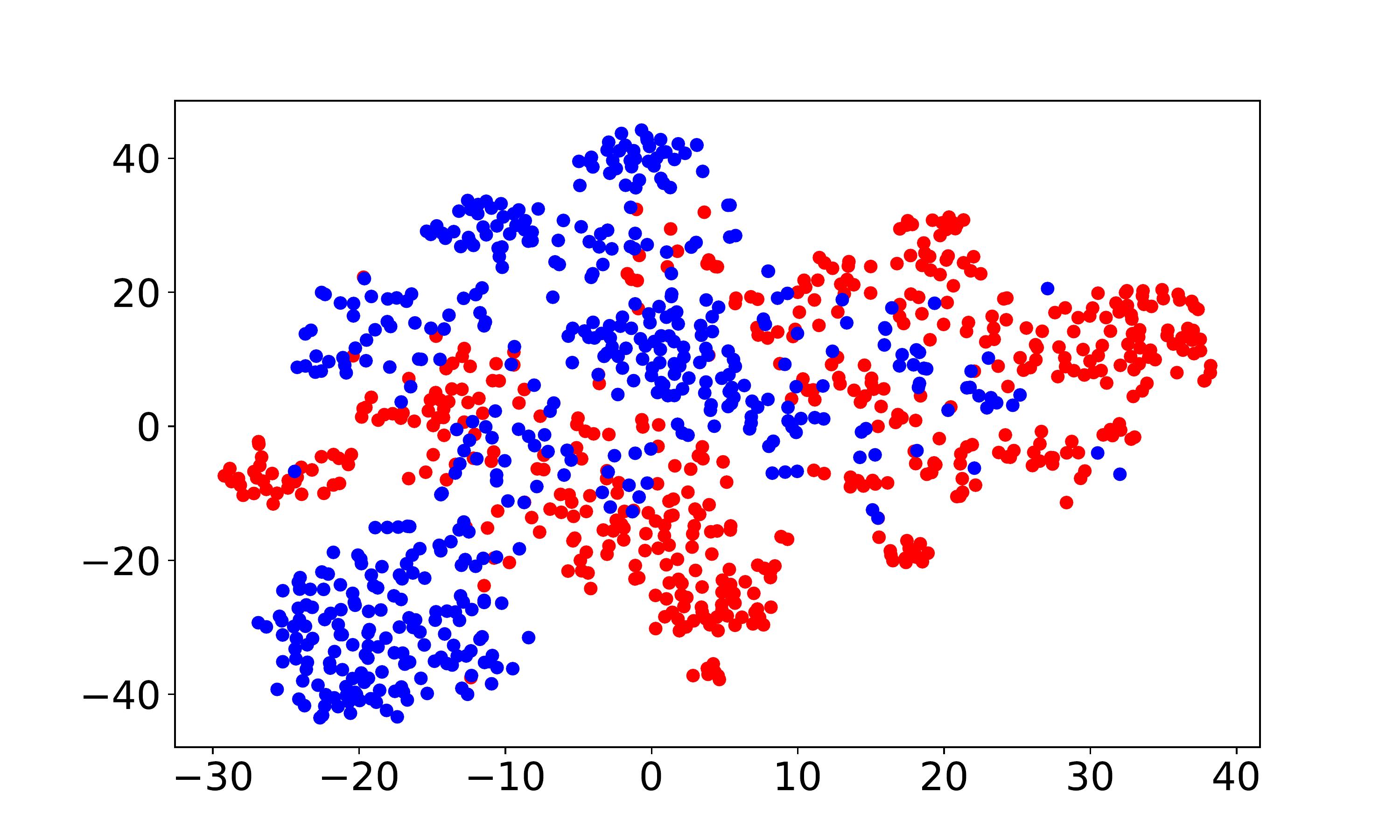}
    \end{minipage}
}
    \subfigure[\newmodelname-Local]{
    \begin{minipage}[t]{0.3\linewidth} 
    \includegraphics[width=5.4cm]{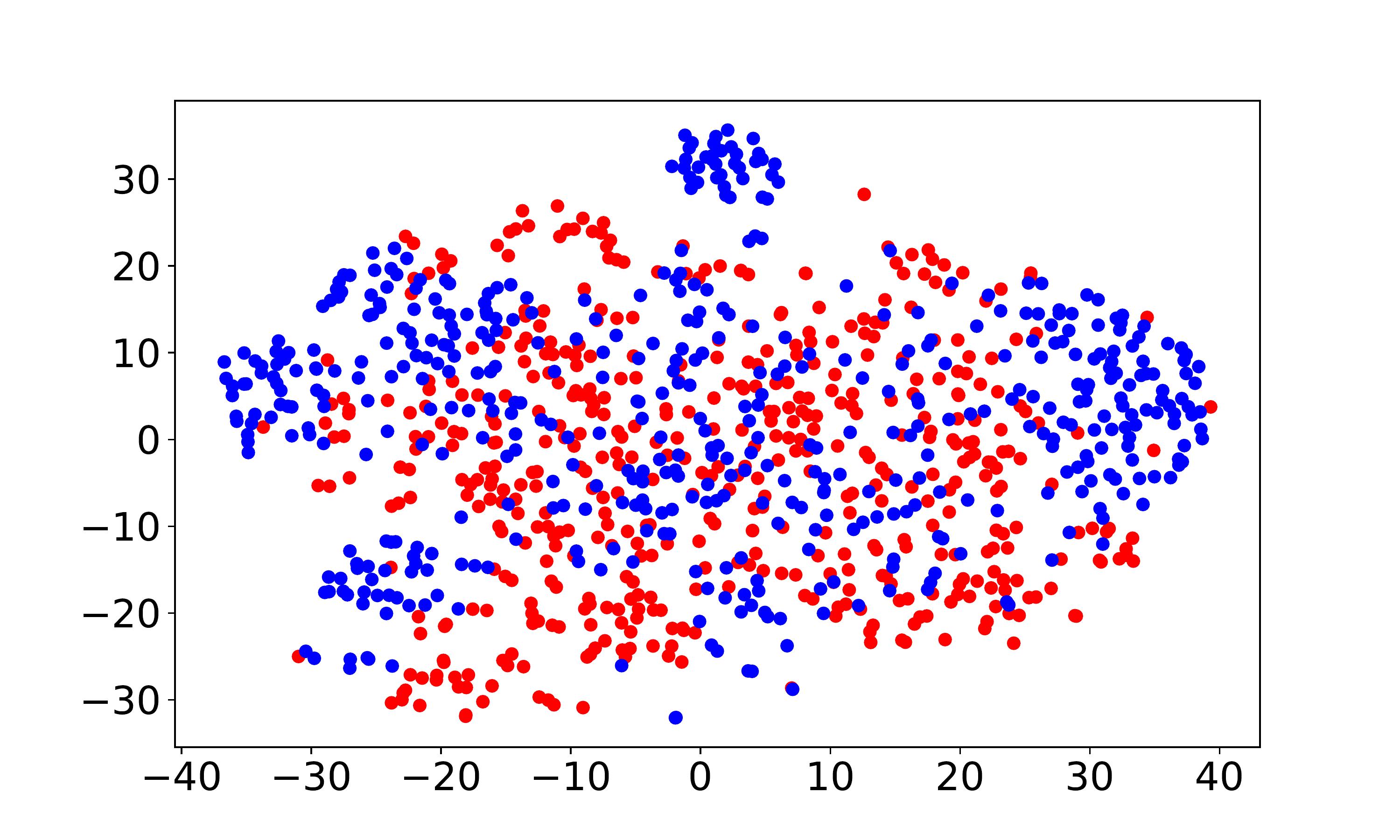}\\
    \includegraphics[width=5.4cm]{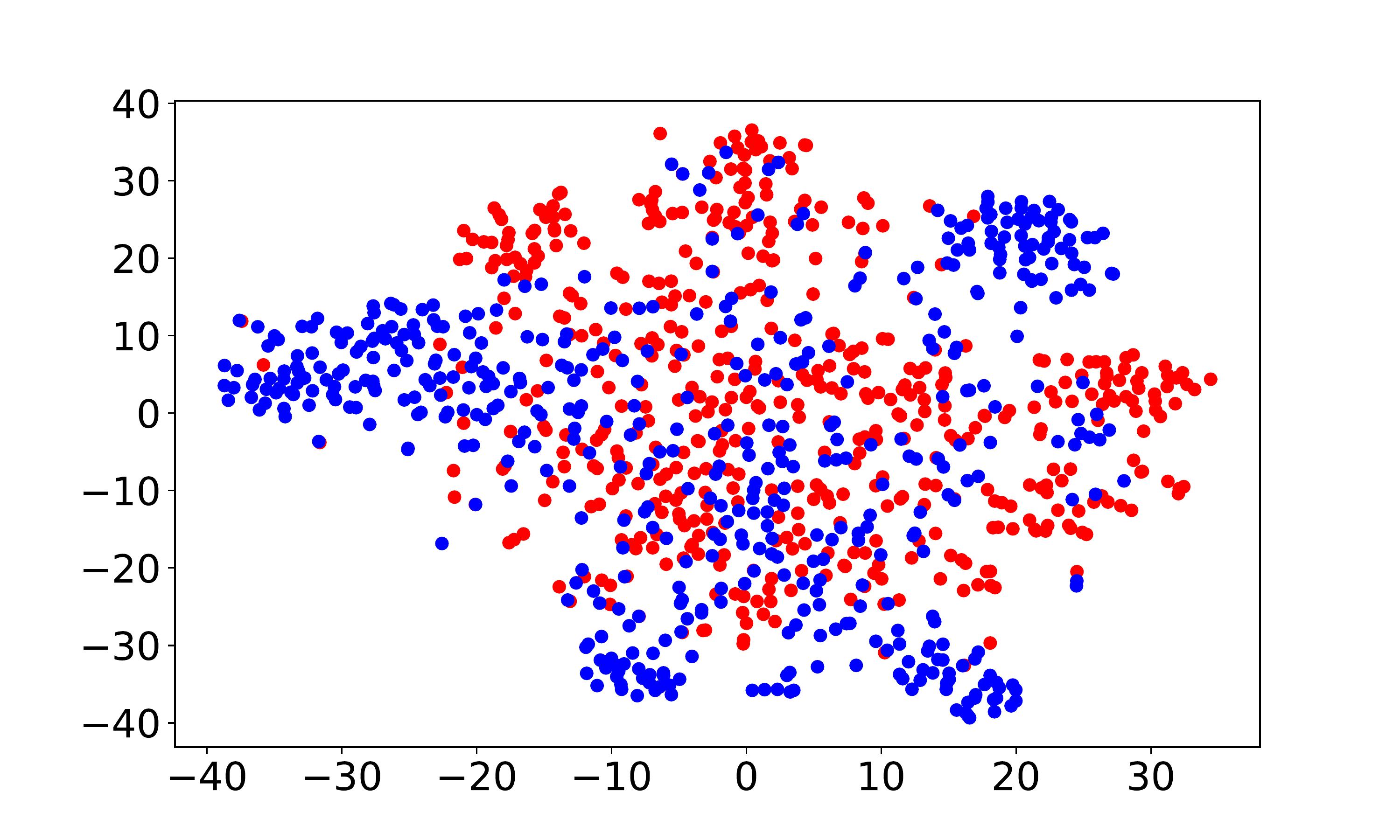}
    \end{minipage}
}
    \subfigure[\newmodelname]{
    \begin{minipage}[t]{0.3\linewidth} 
    \includegraphics[width=5.4cm]{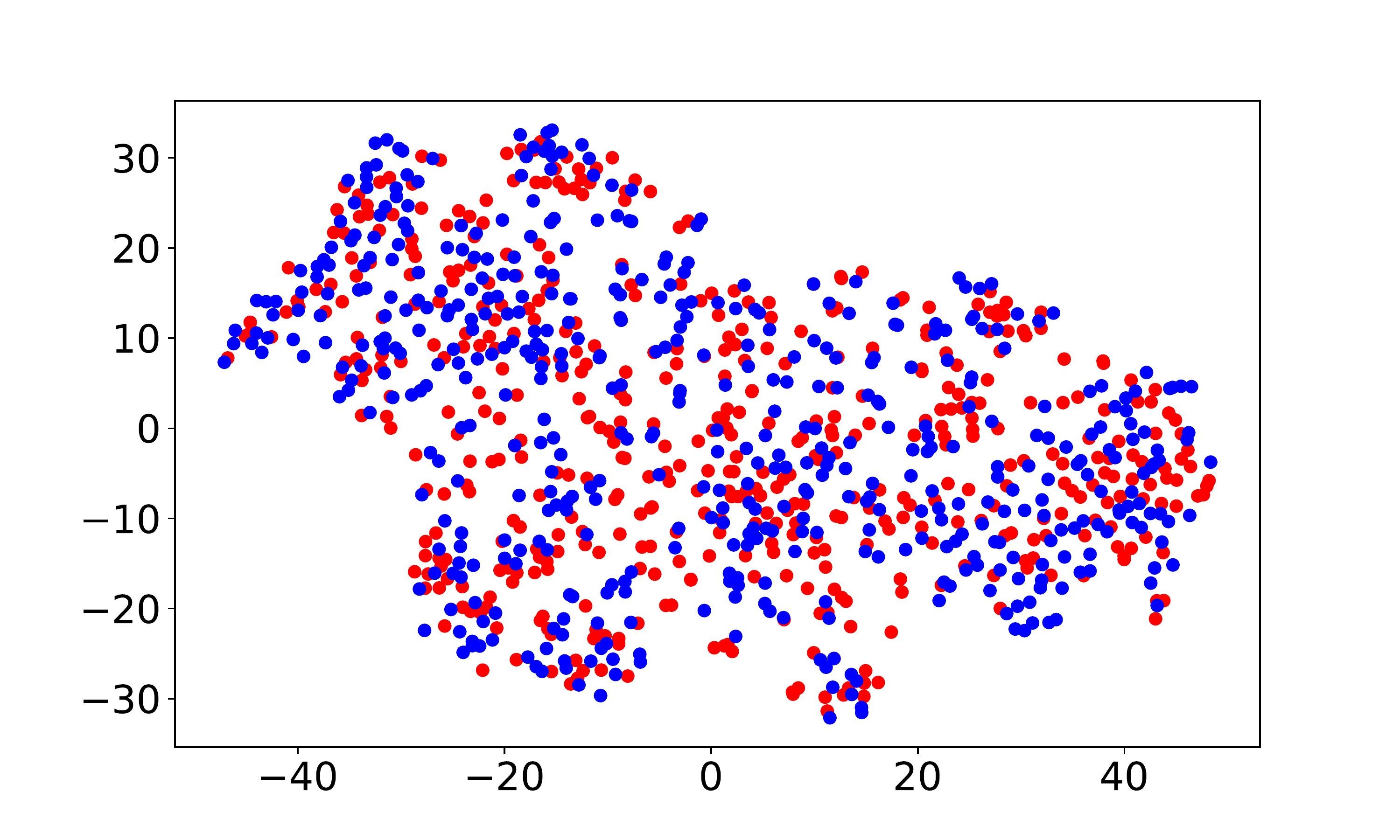}\\
    \includegraphics[width=5.4cm]{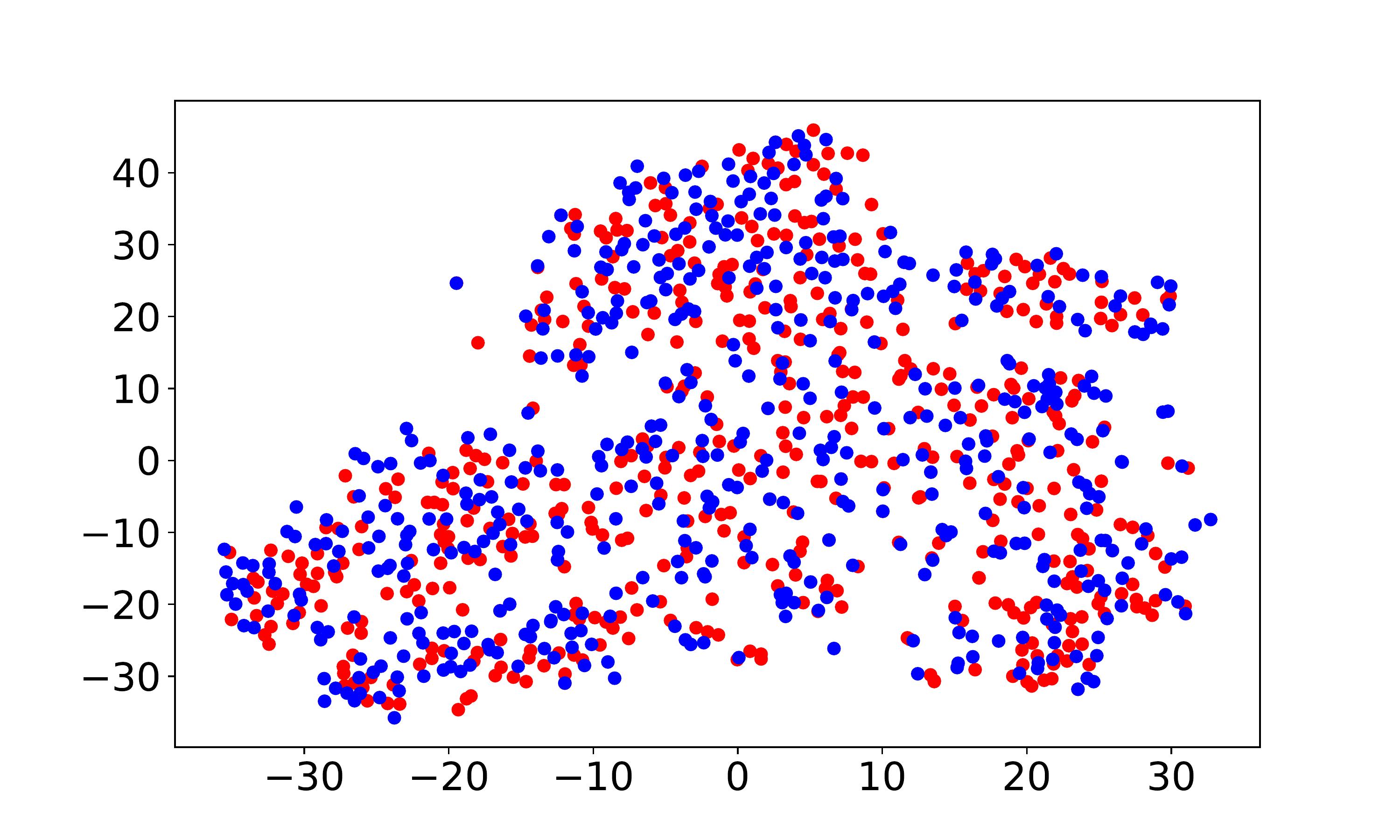}
    \end{minipage}
}
     
	  \caption{The t-SNE visualization of user latent embeddings on \textbf{Amazon Movie} \& \textbf{Book} (first row) and \textbf{Douban Movie} \& \textbf{Book} (second row). The user latent embeddings in the source domain are shown with red dots and those in the target domain are shown with blue dots.}
	  \label{fig:tsne} 
\end{figure*}

\subsection{Recommendation Performance (for RQ1)}
The comparison results on Douban and Amazon datasets are shown in Table 1. 
From them, we can find that: (1) With the decrease of the overlapped user ratio $\mathcal{K}_u$, the performance of all the models will also decrease. 
This phenomenon is reasonable since fewer overlapped users may raise the difficulty of knowledge transfer across different domains.
(2) The conventional shallow model \textbf{KerKT} with multiple separated steps cannot model the complex and nonlinear user-item relationship under the POCDR setting.
(3) Most of the cross domain recommendation models (e.g., \textbf{DML}) perform better than single domain recommendation model (e.g., \textbf{NeuMF}), indicating that cross domain recommendation can integrate more useful information to enhance the model performance, which is always conform to our common sense.
(4) Although \textbf{ETL} and \textbf{DML}  can get reasonable results on conventional cross domain recommendation problem where two domains are highly overlapped, it cannot achieve satisfied solutions on the general POCDR problem. 
When the majority of users are non-overlapped, \textbf{ETL} and \textbf{DML} cannot fully utilize the knowledge transfer from the source domain to the target domain and thus lead to model degradation.
The main reason is that their loss functions only act on the overlapped users without considering the non-overlapped users.
The same phenomenon happens on other baseline models, e.g., \textbf{DARec}. 
(5) \newmodelname~with deep latent embedding clustering can group users with similar behaviors and can further enhance the performance under the POCDR settings.
We also observe that our proposed \newmodelname~can also have great prediction improvement even when the overlapped user ratio $\mathcal{K}_u$ is much smaller (e.g., $\mathcal{K}_u = 30\%$).
Meanwhile our proposed \newmodelname~still obtains better results when the source and target domains are different (e.g., \textbf{Amazon Movie} $\rightarrow$ \textbf{Amazon Clothes}) which indicates the high efficacy of \newmodelname.

\begin{figure*} 
    \centering
    
    \subfigure[Effect of $\lambda_{VL}$]{
    \begin{minipage}[t]{0.23\linewidth} 
    \includegraphics[width=4.1cm]{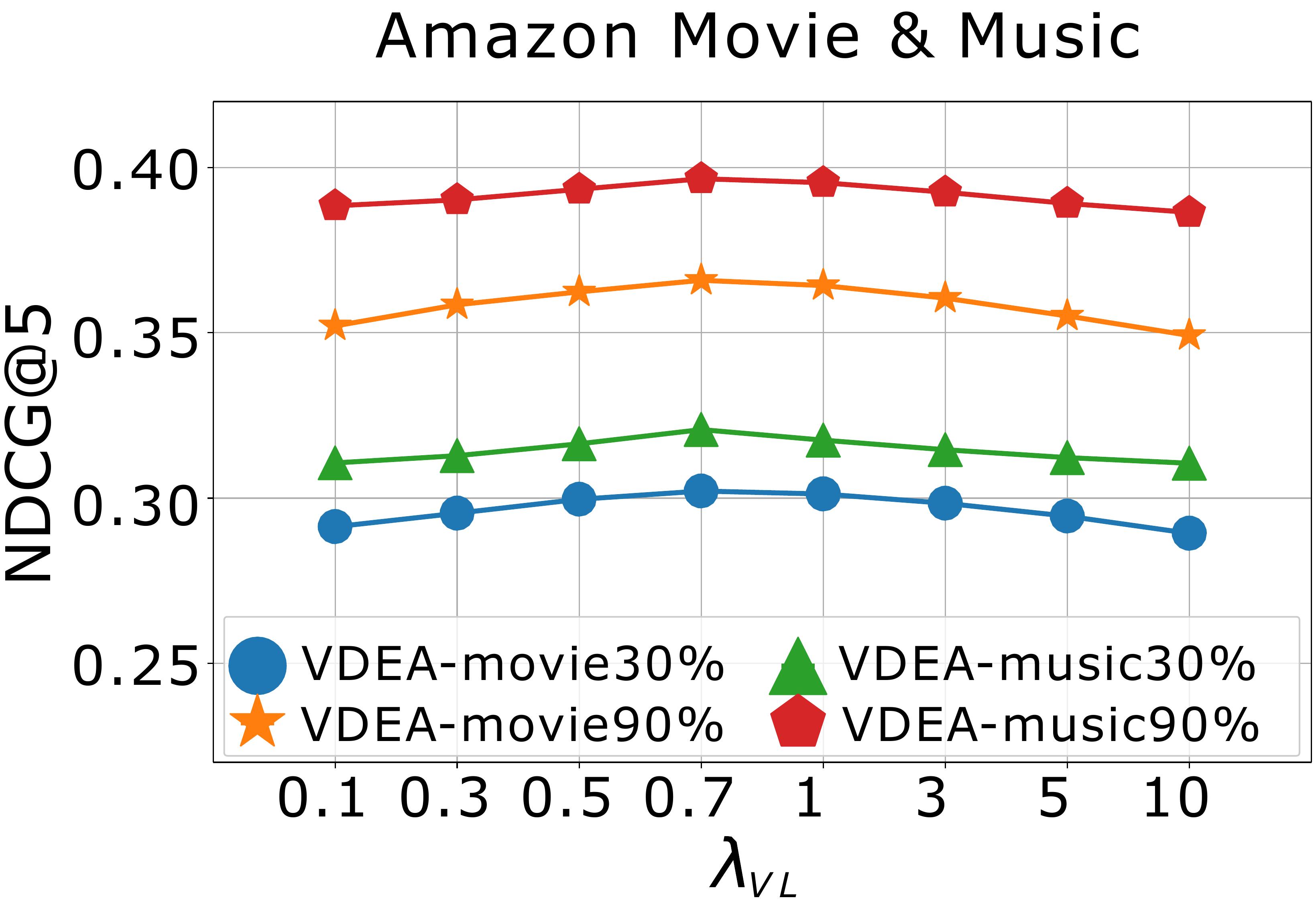}
    \end{minipage}
}
    \subfigure[Effect of $\lambda_{VG}$]{
    \begin{minipage}[t]{0.23\linewidth} 
    \includegraphics[width=4.1cm]{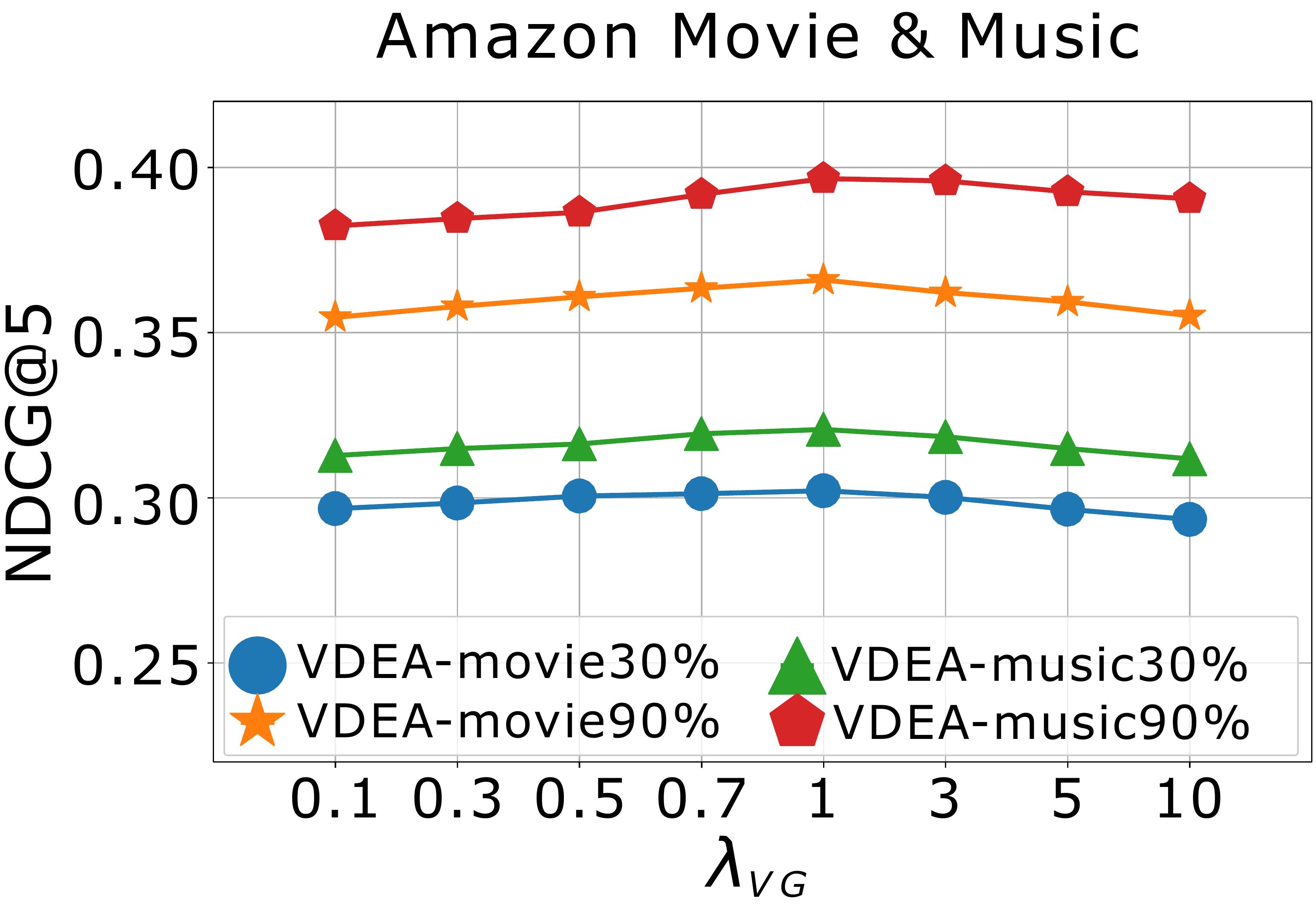}
    \end{minipage}
}
    \subfigure[Effect of $K$]{
    \begin{minipage}[t]{0.23\linewidth} 
    \includegraphics[width=4.1cm]{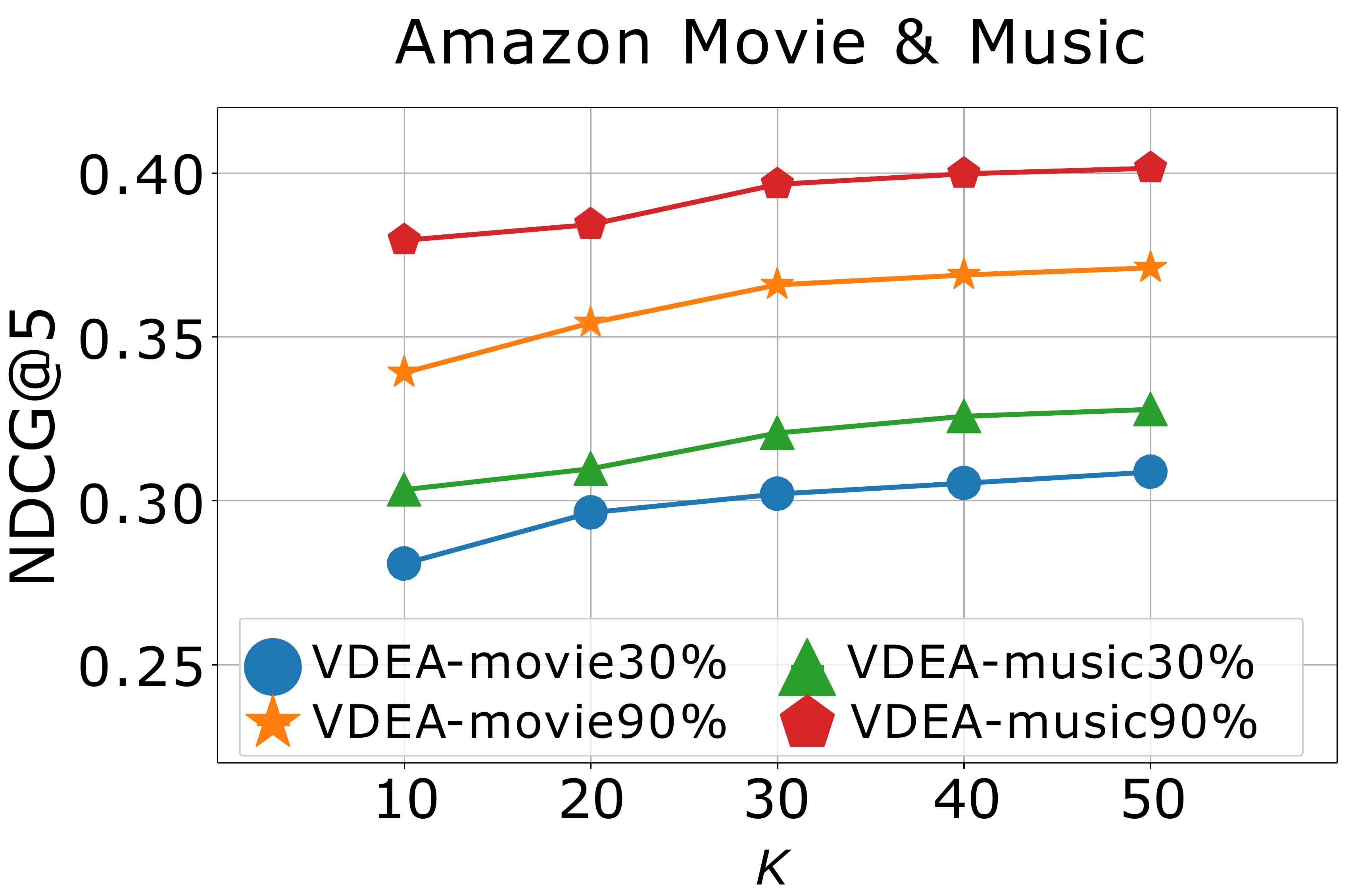}
    \end{minipage}
}
    \subfigure[Effect of $D$]{
    \begin{minipage}[t]{0.23\linewidth} 
    \includegraphics[width=4.1cm]{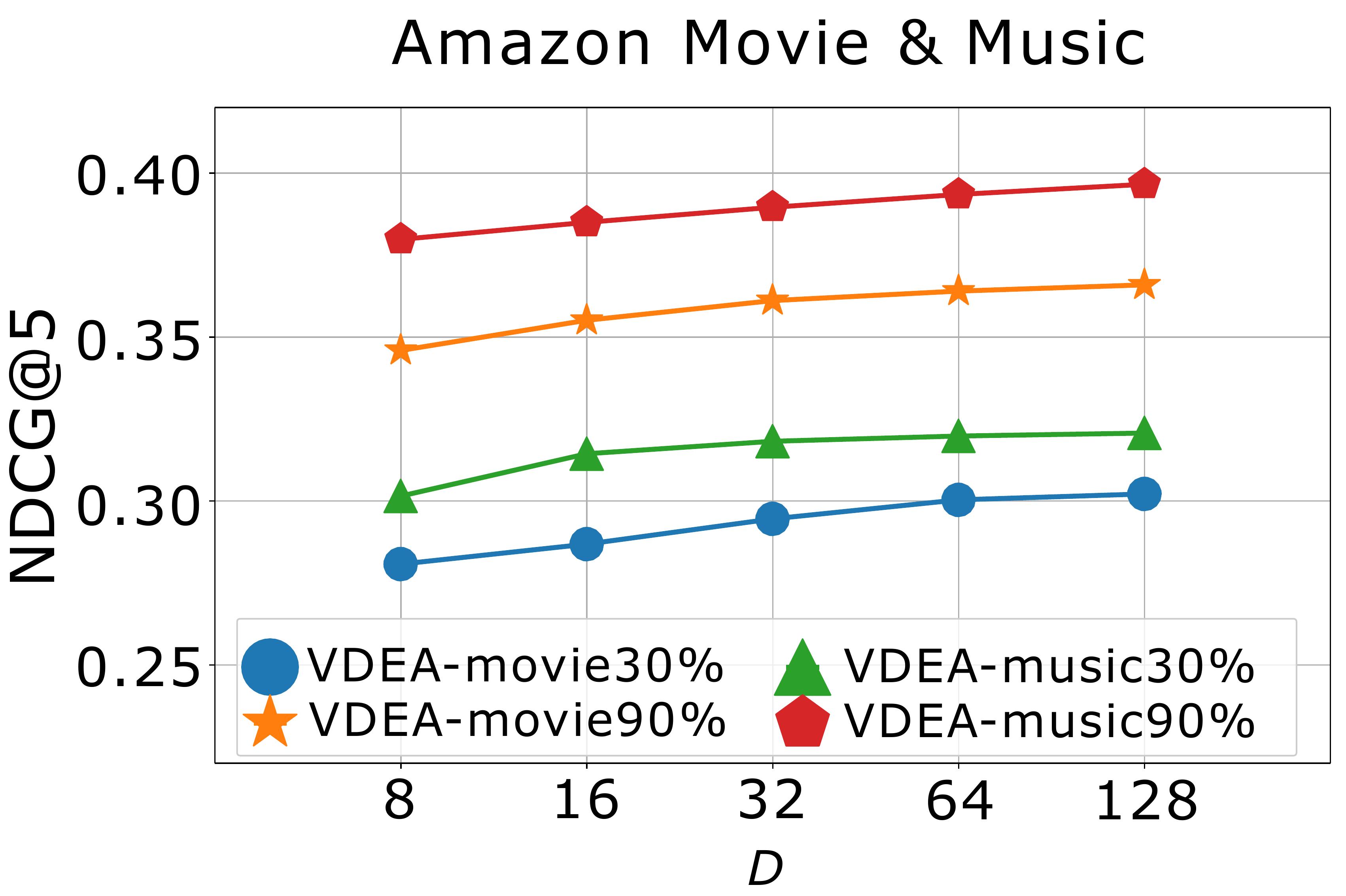}
    \end{minipage}
}
  \vspace{-0.2cm}
	  \caption{(a)-(d) show the effect of  $\lambda_{VL}$, $\lambda_{VG}$, $K$, and $D$ on model performance.}
	  \label{fig:expp} 
\end{figure*}





\subsection{Analysis (for RQ2 and RQ3)}

\nosection{Ablation}
To study how does each module of \textbf{\newmodelname~}contribute to the final performance, we compare \textbf{\newmodelname}~with its several variants, including \textbf{\newmodelname}-Base, \textbf{\newmodelname}-Local, and \textbf{\newmodelname}-Global.
(1) \newmodelname-Base only includes the variational rating reconstruction loss $L_{VR}$, that is $\lambda_{VL} = \lambda_{VG} = 0$.
(2) \textbf{\newmodelname-Local} excludes global embedding alignment for user embedding, that is $\lambda_{VG} = 0$.
(3) \textbf{\newmodelname-Global} adopts local embedding alignment while utilizing the adversarial domain alignment method in \textbf{DARec} for the global embedding alignment.
The comparison results are shown in Table \ref{tab:ablation}.
From it, we can observe that: 
(1) \newmodelname-Base can already achieve good results compared with some other models (e.g., \textbf{MultVAE}), indicating that clustering the users with similar preference inside a single domain can  boost the performance.
However, due to the lacking of knowledge transfer, it cannot exceed the latest cross-domain recommendation model (e.g., \textbf{DML}).
(2) \newmodelname-Local and \newmodelname-Global achieve better results than \newmodelname-Base, showing that adopting the variational embedding alignment module has the positive effect on solving the POCDR problem.
%
%
Aligning the overlapped users can enhance the representation ability for the user embeddings which is also conforms to our intuitions.
However, \newmodelname-Local can only transfer the knowledge on the overlapped users and thus its performance is limited, especially when the overlapped ratio is relatively small.
\newmodelname-Global with the adversarial alignment based on \textbf{DARec} still outperforms than \newmodelname-Local indicates that global distribution alignment is important and essential.
Nevertheless, generative adversarial network with discriminator is unstable and hard to train in practice which hurdles the model performance\cite{dirt-t}.
(3) By complementing both local and global alignment, \newmodelname~can further promote the performance.
Meanwhile, we can conclude that our proposed distribution co-clustering optimal transport method is also much better than traditional distribution alignment (e.g., \newmodelname-Global with \textbf{DARec}).
Overall, the above ablation study demonstrates that our proposed embedding alignment module is effective in solving the POCDR problem.

\nosection{Visualization and discrepancy measurement}
To better show the user embeddings across domains, we visualize the t-SNE embeddings \cite{Laurens2008Visualizing} for \newmodelname-Base, \newmodelname-Local, and \newmodelname.
The results of \textbf{Amazon Movie} \& \textbf{Book} are shown in the first row of Fig.~\ref{fig:tsne}(a)-(c) and the results of \textbf{Douban Movie} \& \textbf{Book} are shown in the second row of Fig.~\ref{fig:tsne}(a)-(c). 
From it, we can see that (1) \newmodelname-Base cannot bridge the user gap across different domains, leading to insufficient knowledge transfer, as shown in Fig.~\ref{fig:tsne}(a).
(2) \newmodelname-Local can align some overlapped users across domains while it cannot fully minimize the global user preference discrepancy, as shown in Fig.~\ref{fig:tsne}(b).
(3) \newmodelname~with local and global embedding distribution alignment can better match and cluster the source and target users with similar characteristics, as shown in Fig.~\ref{fig:tsne}(c). 
The visualization result is consistent with Fig.~\ref{fig:insight}(a)-(c), which illustrates the validity of our model.
Moreover, we adopt $d_{\mathcal{A}}(\mathcal{S},$ $\mathcal{T})=2(1-2\epsilon(h))$ to analysis the distance between two domains, where $\epsilon(h)$ is the generalization error of a linear classifier $h$ which discriminates the source domain $\mathcal{S}$ and the target domain $\mathcal{T}$ \cite{ben2007analysis}. 
Table 4 demonstrates the domain discrepancy on \textbf{Amazon Movie} $\&$ \textbf{Amazon Book} and \textbf{Douban Movie} $\&$ \textbf{Douban Book} tasks using \newmodelname~and several baseline methods.
From it, we can conclude that adopting both local and global alignment in \newmodelname~can better reduce the domain discrepancy between the source and target domains.

\nosection{Effect of hyper-parameters}
We finally study the effects of hyper-parameters on model performance, including $\lambda_{VL}$, $\lambda_{VG}$, $K$, and $D$. 
We first conduct experiments to study the effects of $\lambda_{VL}$ and $\lambda_{VG}$ by varying them in $\{0.1,0.3,0.5,0.7,1,3,5,10\}$ and report the results in Fig.~\ref{fig:expp}(a)-(b).
When $\lambda_{VL},\lambda_{VG} \rightarrow 0$, the alignment loss cannot produce the positive effect.
When $\lambda_{VL}$ and $\lambda_{VG}$ become too large, the alignment loss will suppress the rating reconstruction loss, which also decreases the recommendation results.
Empirically, we choose $\lambda_{VL} = 0.7$ and $\lambda_{VG} = 1.0$.
We then analyse the effect of the cluster number by varying $K = \{10,20,30,40,50\}$ on \textbf{Amazon Music} \& \textbf{Movie} ($\mathcal{K}_u = 30\%, \mathcal{K}_u = 90\%$), and report the result in Fig.~\ref{fig:expp}(c).
We can conclude that more clusters will enhance the model performance but it will naturally consume more time and space as well.
In order to balance the model performance and operation efficiency, we recommend setting $K = 30$ in practice.
%
%
In order to investigate how the latent user embedding dimension affect the model performance, we finally perform experiment by varying dimension $D$ on \newmodelname.
The result on \textbf{Amazon Music} \& \textbf{Movie} ($\mathcal{K}_u = 30\%, \mathcal{K}_u = 90\%$) is shown in Fig.~\ref{fig:expp}(d), where we range $D$ in $\{8,16,32,64,128\}$.
From it, we can see that, the recommendation accuracy of \newmodelname~increases with $D$, which indicates that a larger embedding dimension can provide more accurate latent embeddings for both users and items. 

\begin{table}[t]
\small
\centering
\caption{The results on $d_\mathcal{A}$ for domain discrepancy.}
\label{tab:douban1}
\begin{tabular}{ccc}
\toprule
\multirow{1}{*}{}
& \multicolumn{1}{c}{(Amazon) Movie $\&$ Book}
& \multicolumn{1}{c}{(Douban) Movie $\&$ Book}\\

\midrule
DARec & 1.5032 & 1.5694 \\
DML & 1.4224 & 1.4763 \\
\midrule 
\newmodelname-Base & 1.6510 & 1.6887 \\
\newmodelname-Local & 1.3029 & 1.3516 \\
\newmodelname-Global & 1.2115 & 1.2625\\
\newmodelname & \textbf{1.1644} & \textbf{1.2258} \\
\midrule

\end{tabular}
\vspace{-0.1cm} 
\end{table}

\section{Conclusion}

In this paper, we first propose Variational Dual autoencoder with Domain-invariant Embedding Alignment (\textbf{\newmodelname}), which includes the \textit{variational rating reconstruction module} and the \textit{variational embedding alignment module}, for solving the Partially Overlapped Cross-Domain Recommendation (POCDR) problem.
We innovatively propose local and global embedding alignment for matching the overlapped and non-overlapped users across domains.
We further present Mixture-Of-Gaussian distribution with Gromov-Wasserstein Distribution Co-clustering Optimal Transport (GDOT) to group the users with similar behaviors or preferences.
We also conduct extensive experiments to demonstrate the superior performance of our proposed \newmodelname~models.
In the future, we plan to extend \newmodelname~to item-based POCDR tasks and conduct more comprehensive experiments on new datasets. 
We are also interested in taking side-information into consideration to better solve the POCDR problem. 

\begin{acks}
This work was supported in part by the National Natural Science Foundation of China (No.72192823 and No.62172362) and Leading Expert of ``Ten Thousands Talent Program'' of Zhejiang Province (No.2021R52001).
\end{acks}

\bibliographystyle{ACM-Reference-Format}
\balance
\bibliography{main}

\clearpage

\end{document}